\documentclass[twocolumn]{aastex61}
\usepackage{textcomp}
\usepackage{mathptmx}
\usepackage{amsmath}

\usepackage{amsmath,amsfonts,amssymb}
\usepackage{graphicx}

\received{2018 September 23}
\revised{2018 November 5}
\accepted{2018 November 7}
\submitjournal{PASP}

\begin{document} 
\title{WIRC+Pol: a low-resolution near-infrared spectropolarimeter}

\author{Samaporn Tinyanont}
\affiliation{Department of Astronomy, California Institute of Technology, 1200 E. California Blvd, MC 249-17, Pasadena, CA 91125, USA}
\author{Maxwell A. Millar-Blanchaer} 
\affiliation{Jet Propulsion Laboratory, California Institute of Technology, 4800 Oak Grove Dr, Pasadena, CA 91109, USA}
\affiliation{NASA Hubble Fellow}
\author{Ricky Nilsson} 
\affiliation{Department of Astronomy, California Institute of Technology, 1200 E. California Blvd, MC 249-17, Pasadena, CA 91125, USA}
\author{Dimitri Mawet}
\affiliation{Department of Astronomy, California Institute of Technology, 1200 E. California Blvd, MC 249-17, Pasadena, CA 91125, USA}
\affiliation{Jet Propulsion Laboratory, California Institute of Technology, 4800 Oak Grove Dr, Pasadena, CA 91109, USA}
\author{Heather Knutson}
\affiliation{Division of Geological and Planetary Sciences, California Institute of Technology, 1200 E. California Blvd, Pasadena, CA 91125, USA}
\author{Tiffany Kataria}
\affiliation{Jet Propulsion Laboratory, California Institute of Technology, 4800 Oak Grove Dr, Pasadena, CA 91109, USA}
\author{Gautam Vasisht}
\affiliation{Jet Propulsion Laboratory, California Institute of Technology, 4800 Oak Grove Dr, Pasadena, CA 91109, USA}
\author{Charles Henderson}
\affiliation{Department of Astronomy, Cornell University, 616A Space Science Building, Ithaca, NY 14853, USA}
\author{Keith Matthews}
\affiliation{Department of Astronomy, California Institute of Technology, 1200 E. California Blvd, MC 249-17, Pasadena, CA 91125, USA}
\author{Eugene Serabyn}
\affiliation{Jet Propulsion Laboratory, California Institute of Technology, 4800 Oak Grove Dr, Pasadena, CA 91109, USA}
\author{Jennifer W. Milburn}
\affiliation{Department of Astronomy, California Institute of Technology, 1200 E. California Blvd, MC 249-17, Pasadena, CA 91125, USA}
\author{David Hale}
\affiliation{Department of Astronomy, California Institute of Technology, 1200 E. California Blvd, MC 249-17, Pasadena, CA 91125, USA}
\author{Roger Smith}
\affiliation{Department of Astronomy, California Institute of Technology, 1200 E. California Blvd, MC 249-17, Pasadena, CA 91125, USA}
\author{Shreyas Vissapragada}
\affiliation{Division of Geological and Planetary Sciences, California Institute of Technology, 1200 E. California Blvd, Pasadena, CA 91125, USA}
\author{Louis D. Santos Jr.}
\affiliation{Department of Astronomy, California Institute of Technology, 1200 E. California Blvd, MC 249-17, Pasadena, CA 91125, USA}
\author{Jason Kekas}
\affiliation{ImagineOptix, Cary, NC 27519, USA}
\author{Michael J. Escuti}
\affiliation{ImagineOptix, Cary, NC 27519, USA}
\affiliation{Department of Electrical \& Computer Engineering, North Carolina State University, Raleigh, NC, USA}

\begin{abstract}
    WIRC+Pol is a newly commissioned low-resolution (R$\sim$100), near-infrared (J and H bands) spectropolarimetry mode of the Wide-field InfraRed Camera (WIRC) on the 200-inch Hale Telescope at Palomar Observatory. 
    The instrument utilizes a novel polarimeter design based on a quarter-wave plate and a polarization grating (PG), which provides full linear polarization measurements (Stokes $I$, $Q$, and $U$) in one exposure. 
    The PG also has high transmission across the J and H bands. 
    The instrument is situated at the prime focus of an equatorially mounted telescope. 
    As a result, the system only has one reflection in the light path providing minimal telescope induced polarization.
    A data reduction pipeline has been developed for WIRC+Pol to produce linear polarization measurements from observations. 
    WIRC+Pol has been on-sky since February 2017.
    Results from the first year commissioning data show that the instrument has a high dispersion efficiency as expected from the polarization grating. We demonstrate the polarimetric stability of the instrument with RMS variation at 0.2\% level over 30 minutes for a bright standard star ($J$ = 8.7).
    While the spectral extraction is photon noise limited, polarization calibration between sources remain limited by systematics, likely related to gravity dependent pointing effects. We discuss instrumental systematics we have uncovered in the data, their potential causes, along with calibrations that are necessary to eliminate them.
    We describe a modulator upgrade that will eliminate the slowly varying systematics and provide polarimetric accuracy better than 0.1\%.
\end{abstract}

\keywords{spectropolarimetry, polarization grating}


\section{Introduction}
\label{sec:intro}

The vast majority of astronomical observations are conducted using electromagnetic waves, which have three fundamental properties: intensity, frequency, and polarization. 
Photometry and spectroscopy, which account for most observations in the optical and near-infrared (NIR), are only sensitive to the first two properties of light.
Polarimetry contains information unobtainable just by observing the broadband flux or spectrum of an object. 
Scattering processes, the Zeeman effect near a magnetized source, and synchrotron radiation are among the major astronomical sources of polarized light. 
In particular, scattering-induced polarization can be uniquely used to constrain the geometry of an unresolved scattering region. 
Polarization can reveal asymmetries because in a symmetric scattering region, assuming single scattering, the polarization vector will cancel out when viewed as a point source, leaving no net polarization. 

    WIRC+Pol is a spectropolarimetric upgrade to the Wide-field InfraRed Camera (WIRC; \citealp{wilson2003}), the $8\farcm7 \times 8\farcm7$ NIR (1.1--2.3\,$\mu$m) imaging camera at the prime focus (f/3.3) of the 200-inch Hale telescope at Palomar Observatory, the largest equatorially mounted telescope in the world. 
    WIRC is an opto-mechanically simple, prime-focus, transmissive, in-line centro-symmetrical camera, which has demonstrated an exceptional photometric stability of 100 ppm/30 min, among the best ever recorded from the ground \citep{stefansson2017} .
    Because it is at the prime focus of an equatorially mounted telescope, the light has to reflect only once off of the primary mirror, and the sky does not rotate with respect to the instrument.
    As a result, the instrumental polarization is expected to be low and stable, making WIRC ideal for a polarimetric upgrade.
    The instrument upgrade was motivated by the BD science case summarized below and it has become a part of the observatory's range of facility instruments for other observers in Palomar community.
    The upgrade was enabled by a novel optical device called a polarization grating (PG), that makes a compact and simple low-resolution spectropolarimeter possible. 
    In \textsection \ref{sec:instrument}, we describe the WIRC+Pol instrument including the suite of upgrades we made to the original WIRC instrument. 
    We compare a typical Wollaston prism-based polarimeter (\textsection \ref{sec:polarimeter}) to our PG-based polarimeter (\textsection \ref{sec:pg}).
    The data reduction pipeline is described in \textsection \ref{sec:drp}, and preliminary results exhibiting the instrument's sensitivity are presented in \textsection \ref{sec:results}.
    We discuss possible future instrument upgrades in \textsection \ref{sec:future_upgrade}.
    Conclusions are presented in \textsection \ref{sec:conclusion}.
    
\subsection{Science cases}
    A representative science case for WIRC+Pol and the usefulness of polarimetry is scattering in the atmosphere of brown dwarfs (BDs).
    BDs are substellar objects that cannot sustain hydrogen fusion in their core; hence, they are born hot with heat from gravitational collapse, then radiatively cool as they age. 
    Therefore, their atmospheres progress through a range of temperatures with different chemical processes at play (see a review by \citealp{Kirkpatrick2005}).
    At a narrow temperature range of 1,000--1,200\,K, the atmospheres undergo a sharp photometric and spectroscopic transition.
    The J band brightness increases and the NIR color ($J-K\!s$) turns blue even though the temperature is dropping.
    As brown dwarfs transition from L-type to T-type, spectra start to show broad methane absorption.
    This L/T transition is often explained by a scenario in which clouds of condensates in the L dwarf's atmosphere start to sink below the photosphere, giving way to a clear T dwarf atmosphere.
    While models suggest that observations of T-dwarf atmospheres should be unpolarized, L dwarf atmospheres could be highly polarized due to the scattering of haze and cloud particles \citep{sengupta2009, sengupta2010}. 
    L dwarfs can only be polarized if those scatterers are distributed asymmetrically on the surface, otherwise polarization from different parts of the disk will cancel out. 
    Therefore, a detection of net polarization implies an asymmetry, which can be caused by oblateness of the BD disk due to rotation \citep{marley2011} and/or by patchiness or banding in the cloud distribution \citep{ Stolker2017, dekok2011}.
    While photometry and spectroscopy can provide some constraints on the cloud distribution by observing variability or using the Doppler imaging technique, respectively, they are only sensitive to rotationally asymmetric features.
    Longitudinally symmetric cloud bands like the ones we observe on Jupiter and predicted for brown dwarfs given their fast rotation rates \citep{showman2013}, for example, would go unnoticed from photometric and spectroscopic monitoring.
    Polarimetric observations, therefore, provide a complementary approach: they can further prove the existence of clouds on BDs, cementing their roles in the L/T transition, but then can also reveal the spatial and temporal evolution of these cloud structures. In doing so, polarimetric observations provide important constraints for understanding the atmospheric circulation of brown dwarfs (via general circulation models, GCMs; \citealp{showman2013, Zhang2014, tan2017}).
    Because BD atmospheres bear strong similarities with those of giant gas planets, they provide easily observable proxies to study planetary atmospheres in the high mass regime.

    This science case is only one of many examples where polarimetry is the only method to retrieve spatial information from an unresolved source. Other potential sciences cases of WIRC+Pol include scenarios where scattering occurs in unresolved asymmetric geometries. 
    For example, the study of young stellar objects embedded in their primordial gas and dust cloud, magnetospheric accretion of dust around young ``dipper'' stars, and the ejecta of a core-collapse supernova (CCSN).
    For the CCSN science case, polarimetry is the only way to confirm asymmetry in the explosion mechanism inferred by theoretical models. 
    However, all previous measurements have been conducted in the optical, where light echo from dust in the circumstellar matter (CSM) may mimic the signature of asymmetric ejecta \citep{nagao2017}.
    Multi-wavelength observations, especially in the IR will help distinguishing the source of polarization since CSM dust scattering is inefficient in the IR while electron scattering in the SN ejecta is wavelength independent \citep{nagao2018}. 
    
    Despite polarimeters' unique capabilities, they are not nearly as available and utilized as imagers or spectrographs. 
    This could be partially attributed to the additional complexity of polarimetric instruments, and the fact that most astronomical polarization signals are of an order $<$ 1\%, making them difficult to observe. 
    Furthermore, polarization is not as straightforward to interpret as photometry or spectroscopy. 
    For instance, a 1\% polarization detection from a BD can be caused by inhomogeneity in the cloud coverage, its oblate geometry, a disk around the object, or likely a combination of those sources.   
    Careful radiative transfer modeling is required to meaningfully interpret polarimetric observations. 

    

\section{The Instrument}\label{sec:instrument}
\subsection{A typical polarimeter}\label{sec:polarimeter}
A polarimeter relies on an optical device that differentiates light based on polarization, called an analyzer.
Most designs utilize either a polarizer that transmits only one polarization angle, or a beam-splitting analyzer that splits two orthogonal polarization angles into two outgoing beams.
The polarizer-based polarimeters determine the full linear polarization (i.e.\, Stokes parameters $I$, $Q$, and $U$) by sampling the incoming beam at three, or more, position angles. 
This is typically done either by adding a rotating half-wave plate modulator in front of the analyzer, rotating the whole instrument, or using different polarizers to sample different angles.
An example of an instrument that employs this technique is the polarimetry mode of the Advanced Camera for Surveys on board \textit{Hubble Space Telescope}, which has three polarizers rotated at 60$^\circ$ from each other \citep{debes2016}. 
While polarizers can fit inside a filter wheel of an existing instrument, the polarizer-based design is inefficient because the polarizer blocks about half of the incoming flux and each polarization angle has to be sampled separately.
Alternatively, a polarimeter may use a beam-splitting analyzer, such as a Wollaston prism, that transmits most of the incoming flux into two outgoing beams with minimal loss.
This allows two polarization angles to be sampled simultaneously with one Wollaston prism, and a full linear polarization measurement can be done with only two position angles (though more position angles are typically used to make redundant measurements in order to remove systematics).
This is achieved either with a rotating modulator like in a polarizer-based instrument, or with a split-pupil design with two sets of Wollaston prisms at some angle from each other ({double-wedged Wollaston} \citealp{oliva1997}). 
While being more optically complex, the Wollaston-based design is more efficient than the polarizer-based design because most of the incoming flux gets transmitted to the detector, even though more detector space is needed to image both beams.
As a result, it is more widely used in ground-based instruments, where its higher optical complexity can be accommodated.
There are many polarimeters of this type in use, e.g., the polarimetry and spectropolarimetry modes of the Long slit Intermediate Resolution Infrared Spectrograph (LIRIS; \citealp{manchado2004}) on the 4.2-m William Herschel Telescope.
Both of these polarimeter designs provide only broad-band polarimetry and they have to be coupled with a traditional grating- or grism-based spectrograph to make a spectropolarimeter. 
The end result is an instrument that is large and optically complex.

\subsection{Polarization grating}\label{sec:pg}
WIRC+Pol is a uniquely designed low-resolution spectropolarimeter that can measure linear polarization as a function of wavelength in one exposure, while remaining physically small and optically simple. 
The key to this capability is a compact, liquid crystal polymer-based device called a PG, which acts as a beam-splitting polarimetric analyzer and a spectroscopic grating at the same time \citep{escuti2006, packham2010, millar2014}.
A PG uses a thin polymer film of elongated uniaxially birefringent liquid crystals arranged in a rotating pattern to split an incoming beam based on its polarization into the $m = \pm 1$ diffraction orders while simultaneously dispersing each outgoing beam into spectra \citep[see Figs. 1 and 2 of][]{packham2010}.
A quarter-wave plate (QWP) can be placed before the PG to make a device that splits light based on linear polarization. 
To make this device capable of capturing the full linear polarization in one shot, two halves of the QWP have their fast axis rotated by 45$^\circ$\, and two halves of the PG have the liquid crystals pattern 90$^\circ$\, from each other (see Fig. \ref{fig:schematic} \textit{center}). 
This effectively splits incoming light into four beams with polarization angle 0, 45, 90, and 135$^\circ$. 
In addition, a PG also disperses each beam into a spectrum, with $> 99 \%$ of the incident light into $m = \pm 1$ orders, $\sim 1 \%$ into the $0^{\rm th}$ order and virtually no flux leaking into higher orders.
Moreover, the PG's efficiency is nearly wavelength independent, unlike dispersion gratings which are normally blazed to enhance the efficiency around one specific wavelength.
We demonstrate this property in our transmission measurements in \textsection\ref{sec:throughput}.
These properties make the PG a uniquely efficient disperser and a natural choice for a spectropolarimetric instrument.
Furthermore, a QWP/PG device is thin enough to fit inside an instrument's filter wheel, simplifying its installation in an existing imaging camera.
This is as opposed to a Wollaston prism whose thickness is governed by the required splitting angle. 

\subsection{WIRC Upgrade}\label{sec:upgrade}

\begin{figure*}
	\includegraphics[width = \linewidth]{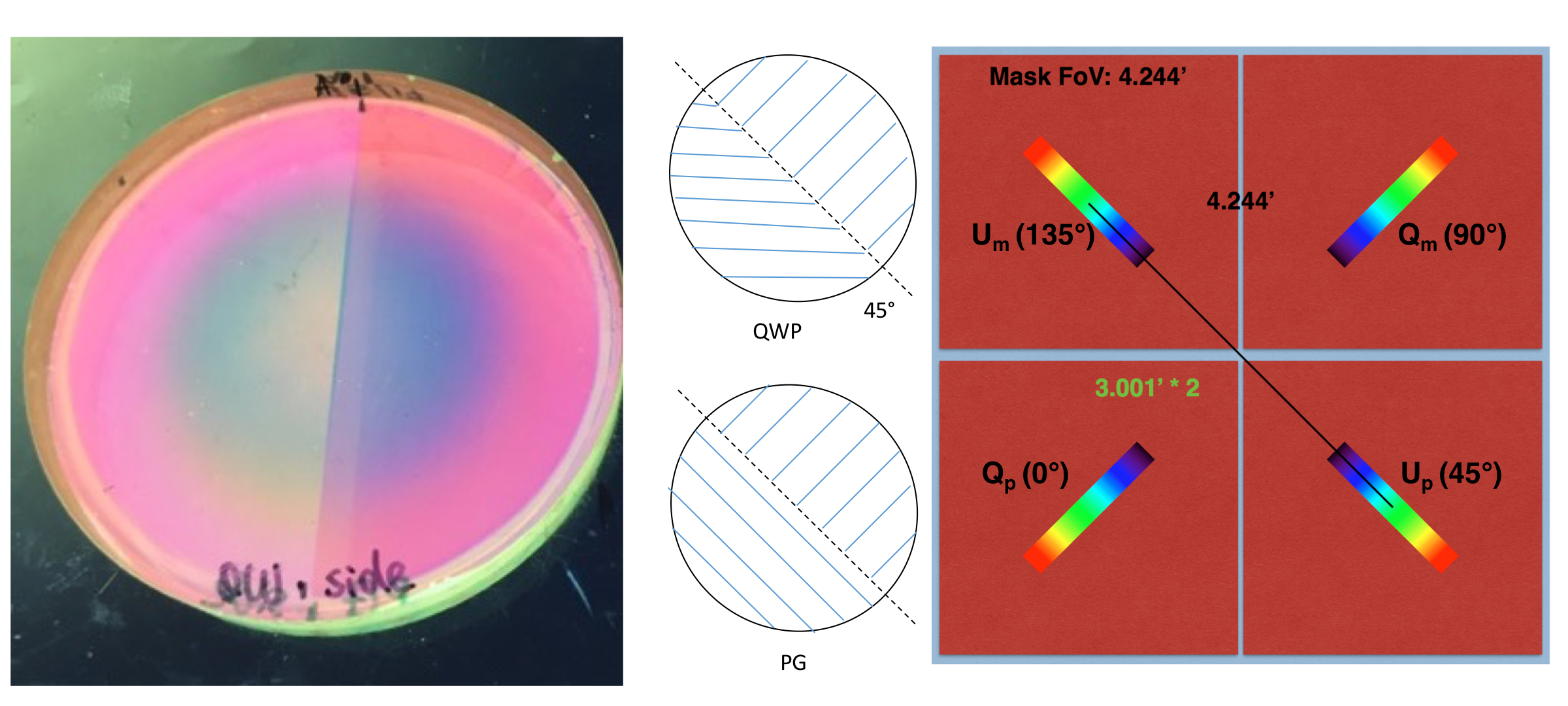}
    \caption{\textit{Left:} Photograph of the actual QWP/PG device installed in WIRC's filter wheel. The line down the middle fo the PG is where the pupil is split. \textit{Center:} Schematics showing the split-pupil design for the QWP and PG. The top figure shows that the QWP's fast axes (notated by the blue lines) are rotated by 45$^\circ$ between the two halves and the bottom shows that the PG's grating axes (also notated by the blue lines) are rotated by 90$^\circ$. As a result, the lower left (upper right) half of the device samples linear polarization angles 0 and 90$^\circ$ (45 and 135$^\circ$). \textit{Right:} Schematic of WIRC+Pol's focal plane image for a single point source. The split-pupil QWP/PG device splits and disperses light into four spectral traces in four quadrants of the detector. Each quadrant is labeled with the corresponding angles of linear polarization. The full field of view (FoV) here is $8\farcm7 \times 8\farcm7$ while the FoV limited by the mask is $4\farcm3 \times 4\farcm3$. The center of each of the four traces in the J band is 3\arcmin\ away from the location of the source in the FoV.}
    \label{fig:schematic}
\end{figure*}

\begin{figure*}
	\includegraphics[width = \linewidth]{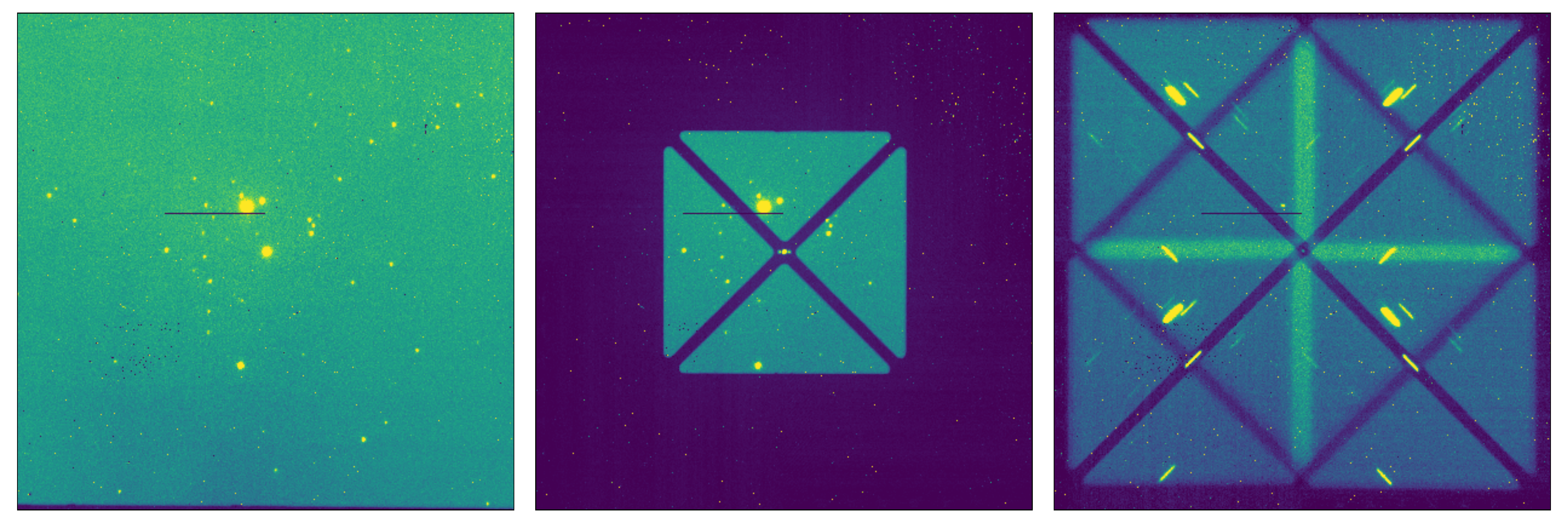}
    \caption{Raw images from WIRC+Pol of the crowded field around HD\,38563, one of the known polarized stars used for calibration, which is the brightest star in this image. Note a bad column running through the star. \textit{Left:} an image from the normal imaging mode with only the J band filter in place. The full field of view (FoV) here is $8\farcm7 \times 8\farcm7$. \textit{Center:} the focal plane mask is put into the optical path at the telescope's focal plane inside WIRC, restricting the field of view to $4\farcm3 \times 4\farcm3$. The metal bars in the center of the field of view hold the three circular holes, each 3\arcsec\ in diameter. \textit{Right:} After the PG is put in place, the field is split into four based on linear polarization, and each of them is dispersed into four quadrants of the detector. The vertical and horizontal bright bars are where the fields overlap. Each point source is dispersed into $R\sim 100$ spectra. Note that the source in the slit has reduced background level. Only the zeroth order (undispersed) image of the brightest star in the field remains easily visible after the PG was inserted.}
    \label{fig:raw_data}
\end{figure*}

For the original WIRC, the converging beam from the telescope primary mirror comes into focus inside of the instrument, then passes through the collimating optics, two filter wheels with a Lyot stop in the middle, and gets refocused onto the detector.
To turn WIRC into a spectropolarimeter, three major components have been installed.

(i) A split-pupil QWP/PG device, manufactured by ImagineOptix \citep{escuti2006}, was installed in the first filter wheel of WIRC, allowing it to be used with the broadband filters J and H, which are in the second filter wheel downstream from the PG in WIRC's optical path.
The initial laboratory testings performed on the Infrared Coronagraphic Testbed \citep{serabyn2016} at the Jet Propulsion Laboratory demonstrated that it responds to a polarized light source as expected.
The device was installed in WIRC in February 2017. 
The filter mount was modified to accommodate the PG, which was installed at 7$^\circ$ angle with respect to the pupil plane to mitigate ghost reflections. 
This filter placement caused some non common path systematic error since outgoing beams from the PG enter the broadband filter (also installed at 7$^\circ$) at different angles, thus seeing different transmission profiles. 
We will discuss this issue in more detail in \textsection\ref{sec:drp} and \ref{sec:throughput}.
The device is optimized for the J and H bands and can potentially be used over the J--H range simultaneously if an additional filter is installed to block the K band thermal emission and limit the sky background.
Laboratory testing confirmed the device's high efficiency, with $<$1\% of total light in the zeroth order image, and over 99\% in the four first order traces, with no leaks into higher orders. 
On-sky tests, to be discussed in \textsection\ref{sec:throughput}, confirmed this measurement. 
The PG is designed with a grating period such that spectral traces on the detector have {1\arcsec} seeing-limited resolution elements of 0.013\,$\mu$m.
This is $R = \lambda/\Delta\lambda \sim$ 100 in the J and H bands.
The QWP/PG is oriented such that the four polarization spectral traces lie on the diagonal of the detector, in order to maximally fill the array, to achieve the largest field of view possible (see Fig.~\ref{fig:schematic} for the schematic and Fig.~\ref{fig:raw_data} for an actual image). 
The large field of view allows for field stars to be used as polarimetric reference to monitor the polarimetric stability.
Fig.~\ref{fig:schematic} \textit{center} shows the QWP's fast axes along with the PG's grating axes. 
The incident light on the lower left (upper right) half of the PG gets sampled at linear polarization angles 0 and 90$^\circ$ (45 and 135$^\circ$) and sent to the lower left and upper right (lower right and upper left) quadrants of the detector (Fig. \ref{fig:schematic}, \textit{right}).
In \textsection\ref{sec:orientation}, we confirmed the orientation of the PG in the instrument by observing the polarized twilight.
We determined that lower left, upper right, lower right, and upper left quadrants correspond to the polarization components with the electric vector at 0, 90, 45, 135$^\circ$ with respect to North, increasing to East, respectively.
Because the 200-inch is on an equatorial mount, these angles remained constant. 
Along with the QWP/PG device, a grism was also installed for a low-resolution spectroscopic mode, WIRC+Spec, for exoplanet transit spectroscopy. 
This observing mode is the topic of an upcoming publication.

(ii) A focal plane mask (Fig. \ref{fig:raw_data} \textit\textit{Center}) was installed at the telescope's focal plane inside the instrument at the same time as the PG was installed. 
The mask restricts the field of view to $4\farcm3 \times 4\farcm3$ so that the field can be split into four quadrants by the PG and still fit into the detector with minimal overlap (see Fig. \ref{fig:raw_data} \textit{center} and {right}). 
The mask can be inserted and removed from the focal plane using a cryogenic motor mechanism. 
The mask has opaque metal bars blocking its two diagonals with three circular holes in the center.
The bars serve to block the sky background emission for a source inside one of the slit holes, providing higher sensitivity.
The holes are {3\arcsec} on-sky in diameter (0.25\,mm at the telescope prime focus), to accommodate the median seeing of {1\farcs2} at Palomar along with the typical guiding error of {1\arcsec}/15 min. 
The mask is made of aluminum and the slit holes have knife-blade edge with a typical thickness of 100 $\rm \mu$m, in order to reduce slit induced polarization, which is proportional to the thickness, and inversely proportional to the width of the slit and the conductivity of the material \citep{keller2001}.
The holes are circular so that any slit-induced polarization is symmetric, and cancel out when the source is centered.
Due to various instrumental systematics uncovered over the course of commissioning, in-slit observations are not yet fully characterized.

(iii) A science-grade HAWAII-2 detector, previously in Keck/OSIRIS \citep{larkin2003}, was installed to replace the engineering-grade detector that had been in place since the failure of the original science-grade detector in 2012.
The engineering-grade device had a defective quadrant that would prevent us from observing four spectra at the same time, and also had many cosmetic defects. 
The existing 4-channel read-out electronics were also upgraded to 32 channels, allowing for a faster read-out time and minimum exposure time of 0.92\,s as opposed to 3.23\,s.
This shorter minimum exposure time enables observations of brighter sources, and proves necessary to access several bright unpolarized and polarized standard stars. 
The detector along with the 32-channel read-out electronics were installed and characterized in January 2017. 
We further discuss these tests in \textsection\ref{sec:detector}.

Along with the hardware upgrades, the instrument's control software received modifications. 
A new control panel was developed to insert and remove the polarimetric mask.
An additional guiding mode based on 2D cross correlation was added to the WIRC guiding script, which previously used to rely on fitting 2D Gaussian profile to stars in the field.\footnote{The 2D cross correlation code was by A. Ginsburg, accessed from \url{https://github.com/keflavich/image_registration}}
With this update, the instrument can now guide on the elongated traces, which is useful both for WIRC+Pol and the spectroscopic mode, WIRC+Spec\footnote{WIRC+Spec is the slitless spectroscopy mode of WIRC installed alongside WIRC+Pol. It involves a low-resolution grism in the filter wheel that work in J, H, and Ks bands with a resolving power of $R \sim 100$.}, especially for faint sources where the zeroth order image of the star is too dim to guide on.
We note here that guiding is done on science images as WIRC has no separate guiding camera.

By adding the focal plane mask, and the beam-splitting and dispersing PG in the optical path, the raw image on the focal plane becomes quite complex. 
Fig.\,\ref{fig:raw_data} shows raw images with (i) just the broadband J filter, (ii) with the focal plane mask inserted, and (iii) with both the mask and the PG inserted. 
From (ii) to (iii), one sees the masked focal plane image split and dispersed into four diagonal directions by the PG. 
Table \ref{tab:specs} summarizes key specifications of WIRC imaging, spectroscopic (WIRC+Spec), and spectropolarimetric (WIRC+Pol) modes. 
Next we describe the data reduction process that turns these complicated images into polarization measurements.

\begin{table}[!t]
    \caption{Specifications of WIRC in different modes.}
    \centering
    \begin{tabular}{|l|l|}
    \hline
        \textbf{Instrument} & WIRC\\
    \hline
        Telescope &Palomar 200-inch Hale\\
        Focus & Prime\\
        Detector & 2048 $\times$ 2048 Hawaii 2\\
    \hline
        \textbf{Spectropolarimetric mode} & WIRC+Pol \\
    \hline
        Bandpass &J, H\\
        Stokes Parameters &I, Q, U (simultaneous)\\
        Spectral resolution & $\simeq 100$ (seeling limited)\\
        Slit size & 3 arcsec \& slitless \\
        Field of View &4.35 $\times$ 4.35 arcmin\\
        Sampling & 0.25 arcsec per pixel\\
        Angular resolution &$\simeq 1".2$ (seeing limited)\\
        Typical p accuracy &1\% \\
    \hline
        \textbf{Spectroscopic mode} & WIRC+Spec  \\
    \hline
        Bandpass &J, H, K\\
        Spectral resolution & $\simeq 100$ (seeing limited)\\
        Slit size & slitless \\
        Field of View &8.7 $\times$ 8.7 arcmin\\
        Sampling & 0.25 arcsec per pixel\\
        Angular resolution &$\simeq 1".2$ (seeing limited)\\
    \hline
         \textbf{Imaging mode} & \\
    \hline
         Wavelength range & 1 to 2.5 microns\\
         Bandpass &BB and NB filters\\
         Field of View &8.7 $\times$ 8.7 arcmin\\
         Sampling & 0.25 arcsec per pixel\\
         Angular resolution &$\simeq 1".2$ (seeing limited)\\
    \hline
    \end{tabular}
    \label{tab:specs}
\end{table}

\section{Data reduction pipeline}\label{sec:drp}
\begin{figure*}
	\includegraphics[width = \linewidth]{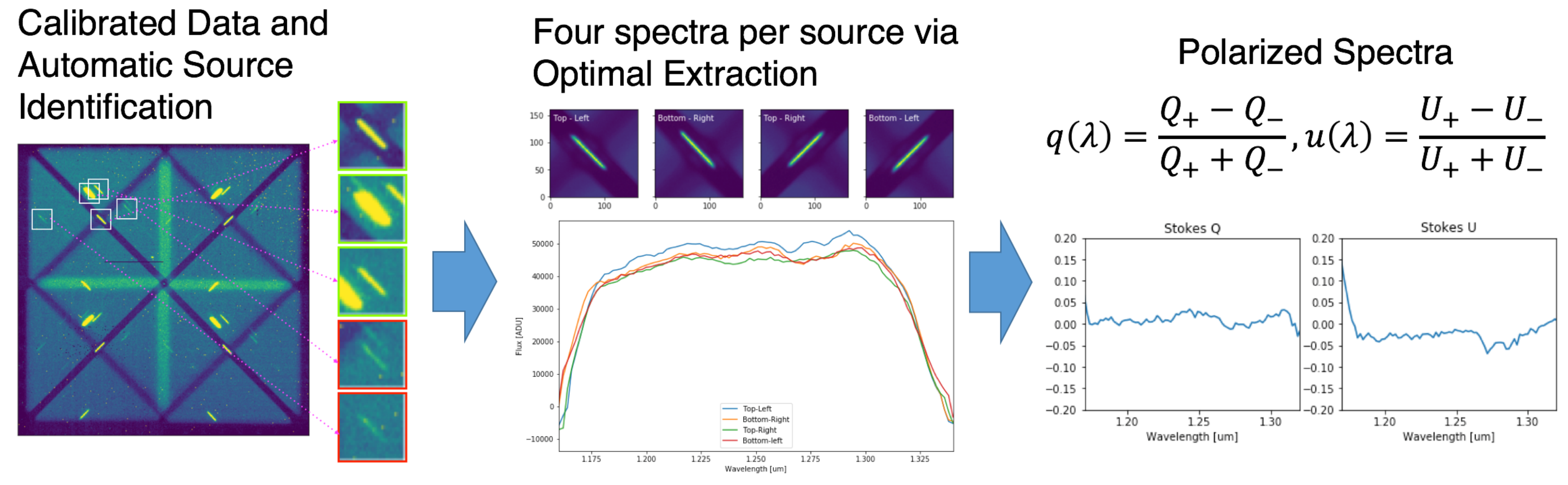}
    \caption{A schematic representing the work flow in the DRP starting with dark subtraction and flat fielding and source identification. Then the DRP extracts four spectra for four linear polarization angles 0, 90, 45, and 135$^\circ$ using optimal extraction. Finally the DRP computes normalized Stokes parameters $q$ and $u$ as functions of wavelengths using the flux spectra from the previous step.}
    \label{fig:DRP}
\end{figure*}

WIRC+Pol is designed for a large survey of hundreds of BDs.
It requires a robust and autonomous data reduction pipeline (DRP) to turn raw observations into polarimetric spectra with minimal user intervention. 
We have developed and tested a Python-based object-oriented DRP that satisfies those requirements.
It is designed with flexibility to be used with future instruments that share WIRC+Pol's optical recipe, i.e.\, split-pupil QWP/PG with four traces imaged at once.
The pipeline is designed to work with the spectroscopy mode, WIRC+Spec, as well.
The schematic of the DRP is shown in Fig. \ref{fig:DRP}. 
Briefly, the DRP first applies standard dark subtraction and flat field correction to raw images. 
It then locates sources in each image, extracts the four spectra for each source, and then computes the polarized spectra. 
To correct for the instrument-induced effects, we normally observe an unpolarized star, chosen from \citet{heiles2000} immediately before or after a science observation.
The DRP is still in constant development, but a working version can be obtained from \url{https://github.com/WIRC-Pol/wirc_drp}.

\subsection{Dark subtraction and flat fielding}
The detector has a measured dark current of approximately 1 e$^-$/s, so dark subtraction is required for long exposures. 
There are a non-negligible number of pixels with high dark current, such that dark subtraction is required even for short exposure time. 
The DRP automatically finds dark frames taken during the night, or nearby nights, and median combines frames with the same exposure times to create master dark frames for each exposure time. 
It then subtracts this master dark frame from science images with the same exposure time. 
In cases when the appropriate master dark with a proper exposure time is not available, the DRP can scale the exposure time of the given dark, although this is not ideal for hot pixel subtraction, and it is generally better to use dark frames with the same exposure time from a different night. 

\begin{figure*}
\centering
\includegraphics[width = 0.8\linewidth]{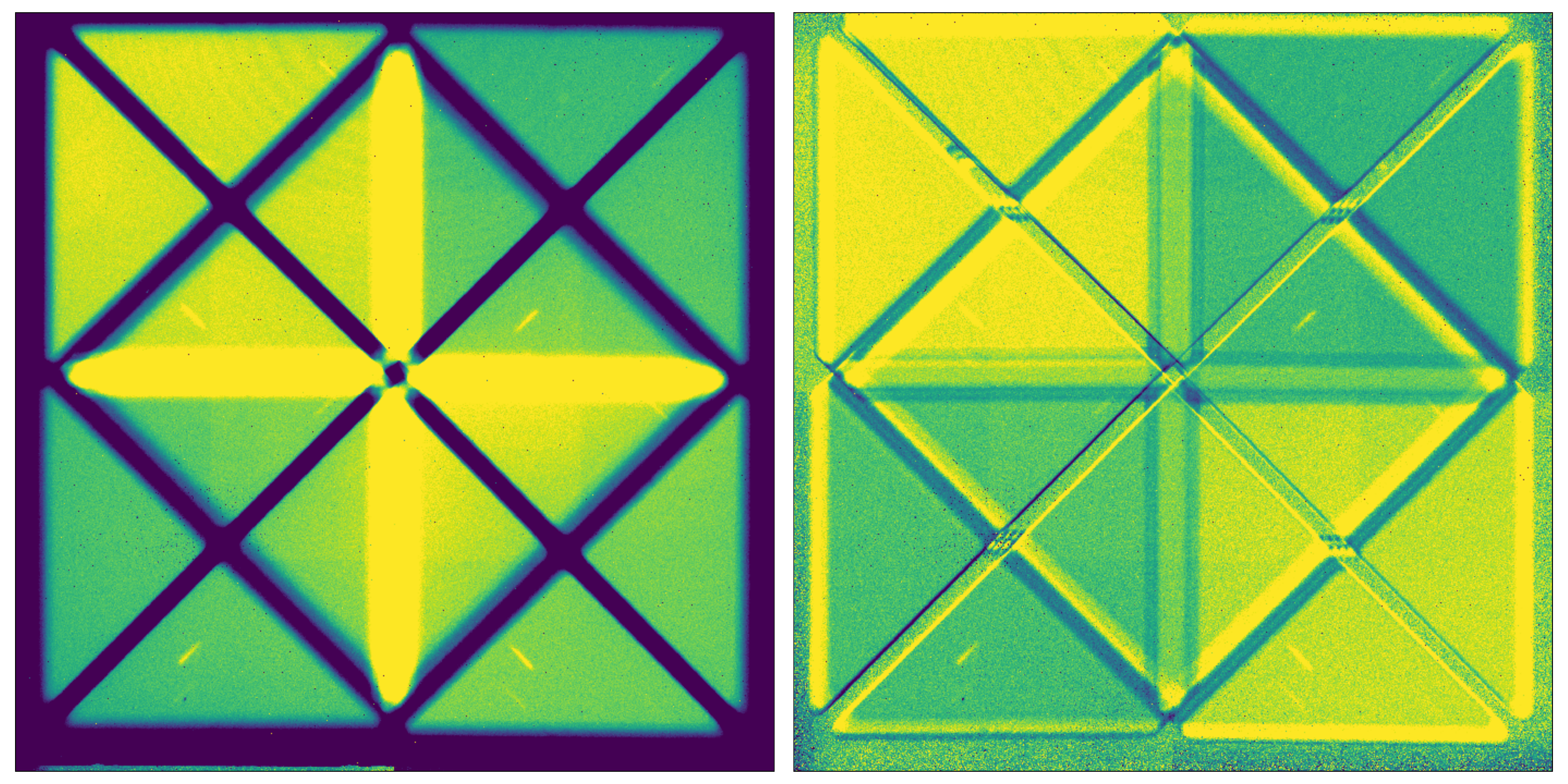}
\caption{A comparison between the same science observation corrected using a flat field without (\textit{left}) and with (\textit{right}) the polarimetric optics (mask and PG) in place. While the correction using the flat without the polarimetric optics does not introduce more artifacts into the image, it fails to correct the uneven illumination due to the polarimetric optics. The leftover flat field variation seen in the \textit{left} figure is removed once we use a flat field image with the polarimetric optics in place. After the flat fielding, one can notice a faint zeroth order background as a rectangle in the center of each image. This contribution is removed during background subtraction. }
\label{fig:compare_flat_fields}
\end{figure*}

Flat field correction is crucial for our observations because we want to compare brightness in four spectral traces far apart on the detector.
An uncorrected illumination variation can cause the four spectral traces to have different flux even when the source itself is unpolarized.
Furthermore, the final polarimetric accuracy depends on the accuracy of this flat field correction. 
Flat fielding is generally difficult for polarimetric instruments due to the fact that one needs an evenly illuminated and unpolarized light source to obtain the calibration. 
As described by \citet{patat2006}, the scenes typically used for flat field correction, such as the twilight sky or a dome lamp, are polarized to some level. 
To circumvent this issue, one may take flat frames without the polarimetric optics in the optical path, which will be agnostic to the source's polarization. 
However, these flat frames will not capture the uneven illumination introduced by the polarization optics, which in our case we found to be significant at the sub-percent level.
We therefore choose to take flat frames with all polarimetric optics in path (the focal plane mask, PG, and the broadband filter). 
We find that the dome flat lamp for the 200-inch telescope is sufficiently unpolarized to provide even illumination in the four quadrants of the detector. 
The spurious polarization introduced here can be subsequently removed by observing an unpolarized standard star.
Fig.~\ref{fig:compare_flat_fields} compares the data corrected by flat fields taken with and without the polarimetric optics on the same scale.
The image corrected by a flat field without the polarimetric optics shows no artifact near the edges of the field of view including the focal plane mask bars.
However, the image corrected by the flat with the polarimetric optics in place shows a much more even background far away from edges. 
This is necessary since the uncorrected background variation is much stronger than the effect from polarization, of order 10\%. 
Another set of dome flats with the PG removed but the mask in place is needed to subtract out the small, additive, zeroth order illumination in the flats with PG. 
This is so that the zeroth order subtracted PG flat represent the PG's efficiency in the $m=\pm1$ only.
We note here that for the flat fielding to not affect the final signal to noise ratio of the spectra to 0.1\% level, the SNR needed is 1000. 
As a result, $10^6$ photoelectrons are needed, and typically the total exposure time of 30 s without PG and 150 s with PG suffice.

\subsubsection{Bad pixel determination}
We identify bad pixels which have peculiar gain in a 3-step process.
First we consider pixels with unusual dark currents.
We use a series of dark exposures, taken during a standard calibration procedure, and compute median and median absolute deviation (MAD) of the count at each pixel.
We choose to use MAD over standard deviation (SD) because the MAD's distribution is close to normal while SD's distribution is not, making it more difficult to make a cut based on the standard deviation of the distribution. 
Since the MAD's distribution is well described by the normal distribution near peak, we use \texttt{Astropy} sigma clipping algorithm to iteratively reject pixels that deviate more than 5$\sigma$ from the mean. 
This creates the first bad pixel map which is particularly sensitive to hot pixels.

Next, we detect dead pixels in flat field images by looking for pixels with spurious values in comparison to their neighboring pixel (local) and to the whole detector (global). 
The local filtering can detect isolated bad pixels well, since their values will be significantly different from the norm established by pixels around them. 
The global filtering, on the other hand, is sensitive to patches of bad pixels where the local filtering fails since these pixels in the center are similar to surrounding, equally bad pixels. 
We note that computing local filtering iteratively can work as well, but may take up more computing time. 
For the local filtering, we use a master flat frame (dark subtracted, median combined, and normalized) obtained each night.
We then create a map of standard deviation, where the value of each pixel is the standard deviation of a box of pixels around it (11$\times$11 box works well).
Pixels that deviate by more than 5$\sigma$ from surrounding pixels are then rejected. 

Finally, for the global filtering, we use the same master flat frame. 
We median filter the master flat to separate the large scale variation component due to the uneven illumination of the focal plane from the pixel-to-pixel variation component.
This step is necessary since the top part of the detector gets up to 20\% more flux, which skews the distribution of pixel response if this large scale variation is not removed. 
The master flat is divided by the large-scale variation map to get an image showing pixel-to-pixel variation.
We find that the pixel-to-pixel map values follow the normal distribution well, so we again use sigma clipping to reject extreme pixels.
At the end of this process, we combine all 3 bad pixel maps: the hot pixels map using dark frames and local and global dead pixels maps from flat frames.
In total, $\sim$ 20,000 pixels are bad, 0.5\% of the whole array.
The time evolution of bad pixels is left for future work.

\subsection{Automatic source detection} 
The spectropolarimetric images obtained by WIRC+Pol contain a composite of four moderately overlapping and spectrally dispersed images of the FOV. Further complexity is introduced by the cross-mask holding the slits/holes in the focal plane. Such an image (see Fig.~\ref{fig:raw_data} \textit{right}) does not lend itself well to most of the standard source detection algorithms provided by, e.g., \emph{Source Extractor} \citep{Bertin1996}, \emph{DAOFIND} \citep{Stetson1987}, or IRAF's \emph{starfind}.\footnote{\url{http://stsdas.stsci.edu/cgi-bin/gethelp.cgi?starfind}} We developed a customized code for automatically detecting source spectra in WIRC+Pol images, which is incorporated into the current pipeline. 

Flat-fielded science images are background subtracted, using a sky image taken $\sim$ 1\arcmin away from the science image to estimate the contribution from sky and mask. As the relative positions of the quadruple of corresponding traces in the four quadrants are known, a single quadrant can be used for source detection. This assumes that the degree of linear polarization of all the sources in the field is small enough not to introduce large differences in brightness between corresponding traces, which is a reasonable assumption for most astrophysical objects. 
Since the four quadrants are just four copies of the same field, we use only the upper left quadrant for source finding.
We convolve the quadrant with a white $J$ or $H$ (depending on the filter in which the science image was obtained) template spectrum that has a FWHM equal to the median seeing at Palomar, and that has the same orientation (assumed to be 45$^\circ$) as the source spectra. This is essentially the traditional `matched filter' method, which effectively enhances the SNR of any image features resembling the template spectrum in a background of white noise. The correlation image is then thresholded, typically at the median pixel value plus $5\sigma$, where $\sigma$ is calculated from background pixels only with sources masked out from the first round of sigma-clipping.
Subsequent masking and labeling of non-zero features gives us a list of positions of detected spectra, ranked by source brightness, and saves user-specified size sub-frames around each spectrum.
Any traces that cross into the regions with dark bars or bright overlapping regions (see Fig. \ref{fig:raw_data} \textit{right}) are rejected.
The corresponding locations of all spectra in the remaining three quadrants are then calculated, and all sub-frames containing `good' spectra are passed on to the spectral extraction part of the pipeline.

\subsection{Spectral extraction}
The spectral extraction step employs a classical optimal extraction algorithm by \citet{Horne1986}.
For each sub-frame of a spectral trace, we first have to estimate (i) the variance for each pixel and (ii) the sky background.
For the dark subtracted, flat field corrected, and data $D$ in the data number ($\rm ADU$) unit, we obtain the variance image estimate by 
\begin{equation} \label{eq:var_init}
V = \sigma^2_{RN}/Q^2 + D/Q
\end{equation}
where $\sigma_{RN}$ is the read-out noise RMS in the electron unit and $Q$ is the gain in $\rm e^-/ADU$ (12 $\rm e^-$ and 1.2 $\rm e^-/ADU$ respectively for WIRC+Pol, see \textsection\ref{sec:detector}).
To estimate sky background, $S$, we fit a 2D low order polynomial (default to second order, but it is user adjustable) to the image which has the spectral trace masked out.

This optimal extraction algorithm requires the spectral trace to be aligned with the detector grid, which is not the case for WIRC+Pol data.
Therefore, we first rotate $D$, $V$, and $S$ images using the \texttt{warpAffine} function from \texttt{OpenCV} with a rotation matrix given by the \texttt{getRotationMatrix2D} function.
We measured the angle to rotate by fitting a line to the brightest pixel in each column of the thumbnail $D$ and we rotate around the center of the thumbnail.
Next we describe the extraction algorithm.
In a standard, non-optimal, spectral extraction procedure, the flux and variance at each wavelength bin is determined by the sum of the background subtracted data along the spatial direction in that wavelength bin. This can be written as 
\begin{align} \label{eq:fstd}
F_{\lambda,\mathrm{std}} &= \Sigma_x \, (D_{\lambda,x}-S_{\lambda,x}) \\ \label{eq:vstd}
\sigma^2_{F_{\lambda,\mathrm{std}}} &= \Sigma_x \, V_{\lambda, x}
\end{align}
The summation boundary in the spatial ($x$) direction is $\pm 9 \sigma$ from the peak of the trace where $\sigma$ is determined by fitting a Gaussian profile along the spatial direction of the brightest part of the trace. 
This extraction method is non-optimal because it gives equal weight to the noisy wings of the spectral trace as it does the peak.
As a result, the extracted 1D spectra are noisier, especially for low SNR data.

The optimal extraction algorithm solves this issue by fitting an empirical spectral profile to the trace and assigning more weight to the less noisy region.
The key to this optimization is the profile image, $P$, of the data, which represent the probability of finding photons in each wavelength column as a function of spatial row. 
The profile image can be constructed as follows:
(i) For each wavelength column $\lambda$ of $D-S$, divide each pixel by $F_{\lambda,\rm std}$ from \eqref{eq:fstd}.
This gives us a normalized flux in each column.
(ii) We assume that the profile varies slowly as a function of $\lambda$. 
As such, we can smooth $P$ by applying a median filter in the $\lambda$ direction, with the default filter size of 10 pixels. 
(iii) Then for each column ($\lambda$), we set all pixels with negative $P$ to 0, and normalize $P$ such that $\Sigma_x P_{\lambda,x} = 1$

With the knowledge of the spectral profile, we can revise the variance estimate from \eqref{eq:var_init} by 
\begin{equation} \label{eq:var_revise}
    V_{\rm revised} = \sigma^2_{\rm RN}/Q^2 + |FP+S|/Q
\end{equation}
where we replace the noisy data $D$ by a model based on the measured flux $F$ and profile $P$.
(Note that $FP$ term is $F_{\lambda, \rm std}$ from \eqref{eq:fstd} multiplying the image $P$ column-by-column).
Bad pixels that are not captured earlier in the calibration process and cosmic ray hits can be rejected by comparing the data to the model:
\begin{equation}
    M = (D - S - FP)^2 < \sigma^2_{\rm clip}  V_{\rm revised}
\end{equation}
where $M$ is 1 where the difference is within some $\sigma_{\rm clip}$ of the expected standard deviation.
At this stage, we can optimize the flux and variance spectra by
\begin{align} \label{eq:f_opt}
F_{\lambda, \rm opt} &= \frac{\Sigma_x MP(D-S)/V}{\Sigma_x MP^2/V} \\ \label{eq:var_opt}
V_{\lambda, \rm opt}  &= \frac{\Sigma_x MP}{\Sigma_x MP^2/V}
\end{align}
If needed, one can iterate this process by reconstructing the profile image using this new optimized flux, then repeat the following steps (eq. \eqref{eq:var_revise} to \eqref{eq:f_opt}) to arrive at a cleaner final optimized flux and variance.
This spectral extraction process is to be run on four spectral traces for each source. 
Adopting the Stokes parameters formulation of polarization, we call the traces corresponding to 0, 90, 45, and 135$^\circ$ respectively $Q_p$, $Q_m$, $U_p$, and $U_m$.
The detector locations of these traces are lower left, upper right, lower right, and upper left (see Fig. \ref{fig:schematic} \textit{right}).

\subsection{Wavelength solution}
For the polarimetric calculation in the next step, it is crucial to ensure that all spectra are well aligned in wavelength. 
A precise absolute wavelength solution is not necessary at this step, so we first compute a relative wavelength solution between the four spectral traces.
Aligning four spectra in wavelength is complicated because WIRC+Pol's filters and PG are tilted at 7$^\circ$ away from being orthogonal to the optical axis.
As a result, the filter transmission profile differs for the four traces since the outgoing beams from the PG hit the filter at different angles \citep{ghinassi2002}.
This effect is also field dependent since a source observed at different positions on the detector enter the filter at different angles.
As a result of this profile shift, we cannot rely on the filter cutoff wavelengths to compute the wavelength solution.
The best practice is to first align all the high SNR spectra (SNR$\sim$1,000 per spectral channel) of a standard star to each other, relying on atmospheric absorption features at 1.26-1.27 $\mu$m due to $\rm O_2$ in the J band and multiple $\rm CO_2$ lines in the H band. 
These features can be seen clearly in the absolute throughput plot shown in Fig.~\ref{fig:throughput}. 
We note that some standard stars also have the hydrogen Paschen-$\rm \beta$ line at 1.28 $\mu$m and multiple Brackett lines in the H band that we can use for alignment as well.
Currently we align the trough of the absorption line manually. 
After the four spectra of the standard star are aligned in wavelength, we can align spectra of our source to the corresponding spectra of the standard star. 
It is important that the source and the standard are observed at a similar position on the detector, so that the filter transmission profile for the two are identical. 
We found that the guiding script (described at the end of \textsection\ref{sec:upgrade}) can reliably put a new source on top of a given reference star to within a pixel.
We then can rely on the filter transmission cutoffs to align each of four traces of the source to those of the standard. 
For the absolute wavelength solution, we assume that wavelength is a linear function of the pixel position, which is reasonable at this low spectral resolution. 
The spectral dispersion in $\mu$m per pixel is given by comparing the measured spectrum (in pixels) to the filter transmission profile.
The wavelength zeropoint is calibrated to the atmospheric absorption features used for alignment.

\subsection{Polarization calibration and computation}\label{sec:pol_computation}
Linear polarization Stokes parameters ($q$ and $u$) are the normalized flux differences between the two orthogonal pairs.
\begin{align}
    q &= (Q_p - Q_m)/(Q_p + Q_m) \\ 
    u &= (U_p - U_m)/(U_p + U_m) \
\end{align}
The degree and angle of linear polarization can be computed with following equations:
\begin{align}
        p &= \sqrt{q^2 + u^2} \\
    \Theta &= 0.5 \tan^{-1} (u/q)
\end{align}
In practice, however, the calculation is complicated by non common path effects in WIRC+Pol's optical path.
Firstly, the camera has an uneven illumination across the field of view---typical of a wide field instrument.
This can introduce a flux difference between e.g. $Q_p$ and $Q_m$ when the source is unpolarized.
This effect remains at some level even after a flat field correction.
Secondly, as mentioned earlier, the PG and all filters in WIRC were installed at 7$^\circ$ with respect to perpendicular of the optical axis to mitigate ghost reflections.
As a result, the upper and lower spectral traces enter the broadband filters (either J or H) downstream from the PG at different angles, and experience slightly different filter transmission profiles \citep{ghinassi2002}.
This shift can be seen in the transmission curves shown in Fig.~\ref{fig:throughput} (to be discussed in more details in \textsection\ref{sec:throughput}).

In order to remove these non-common path effects, we follow the calibration scheme described here.
For brevity, we consider the $Q$ pair, as the process for the $U$ pair is identical.
First, we observe an unpolarized standard star at the same detector position as our target. 
The intrinsic spectrum of this standard is $S(\lambda)$, which is the same for all four traces since the standard is not polarized.
We have the observed spectrum
\begin{align}
    S'_p &= S(\lambda)\, A_1(\lambda)\, F_p(\lambda) \\
    S'_m &= S(\lambda)\, A_1(\lambda)\, F_m(\lambda)
\end{align}
where $F_{p,m}(\lambda)$ are the filter transmission functions seen by the plus (lower) and minus (upper) traces.
Note here that the filter transmission function depends on the angle of incidence on the broadband filter, therefore it also changes across the field of view. 
$A_1(\lambda)$ is the other transmission function which is similar for both traces (e.g., atmosphere, telescope reflective coating, etc.)
If our science target has intrinsic fluxes $I_p$ and $I_m$ due to some intrinsic polarization, we will observe
\begin{align}
    I'_p &= I_p(\lambda)\, A_2(\lambda)\, F_p(\lambda) \\
    I'_m &= I_m(\lambda)\, A_2(\lambda)\, F_m(\lambda)
\end{align}
where $A$ may change due to e.g. changing atmosphere. 
Recall that if this source has an intrinsic normalized Stokes parameter $q$, then $q = (I_p - I_m)/(I_p + I_m)$.
We remove the transmission functions by dividing the observed target spectrum by the observed standard spectrum which sees the same filter transmission profile $F$.
The ratio $A_1/A_2$ will cancel out here as well. 
We can then recover this intrinsic polarization by computing
\begin{equation}\label{eq:qu}
        \frac{{I_p'}/{S_p'} -{I_m'}/{S_m'}}{{I_p'}/{S_p'} +{I_m'}/{S_m'}} = \frac{I_p - I_m}{I_p + I_m} = q
\end{equation}
Note that the standard star intrinsic spectrum term $S(\lambda)$ cancel out because it is the same for all 4 traces.
A similar process can be applied to the $U$ pair to measure $u$ as well.

For polarimetric uncertainties, we first obtain uncertainties of the measured spectrum by computing the standard deviations in each spectral bin for each source and standard spectrum from the series of exposures. 
Then we compute uncertainties of the flux ratios $I/S$ by error propagation assuming normal distribution.
Let's denote flux ratios in \eqref{eq:qu} by $Q_p = I'_{Qp}/S'_{Qp}$ and so on. 
The uncertainties to $q$ and $u$ are also calculated by error propagation, assuming Gaussian error, using the following equations:
\begin{align}
    \sigma_q &= \frac{2}{(Q_p + Q_m)^2} \sqrt{(Q_m\sigma_{Q_p})^2 +(Q_p\sigma_{Q_m})^2} \\ 
    \sigma_u &= \frac{2}{(U_p + U_m)^2} \sqrt{(U_m\sigma_{U_p})^2 +(U_p\sigma_{U_m})^2} \\
    \sigma_p &= \frac{1}{p} \sqrt{ (q \sigma_q)^2 + (u \sigma_u)^2} \\
    \sigma_\Theta &= \frac{1}{2p^2} \sqrt{(u \sigma_q)^2 + (q \sigma_u)^2}
\end{align}
We have confirmed from the commissioning data that $q$ and $u$ follow normal distribution.
However, $p$ is a non-negative quantity following a Rice distribution with a long positive tail \citep{jensen2016}.
Its mean value is biased to the positive and has to be corrected, especially when the value is close to zero, using \citep{wardle1974}
\begin{equation}
	p* = \sqrt{p^2 - \sigma^2_{p}}
\end{equation}



\section{Instrument Commissioning}\label{sec:results}

\subsection{Detector characterization}\label{sec:detector}

\subsubsection{Linearity and dark current measurement}
Infrared detectors have a linear response to photon counts up to a certain amount. 
We measure this linearity limit by taking flat exposures at different exposure times and plot the mean count as a function of exposure time. 
To quantify the linearity, we fitted a line through the first few data points where the response is still unambiguously linear. The deviation from this fit is then the degree of non-linearity. 
We found that the new H2 detector is linear to 0.2\% level up to 20,000 $\rm ADU$ and to 1\% level at 33,000 $\rm ADU$.

The dark current can be measured by taking dark exposures at various exposure times and fitting a linear relation to the median count. 
We measure the median dark current across the detector to be 1 $\rm e^-/s$.
We note here that WIRC does not have a shutter, and dark frames are obtained by combining two filters with no overlapping bandpass, typically Brackett-$\gamma$ and J band filters. 

\subsubsection{Gain and read-out noise}
We measure the gain and the read-out noise of the detector using the property of Poisson statistics where the variance equals the mean value. 
If $N$ is the number of photoelectrons detected and $\rm ADU$ is the measured count, we have that $N = g \rm ADU$ where $g$ is the gain factor in $\rm e^-/ADU$.
The variance of the count is a sum of the photon shot noise and the detector read-out noise: $g^2 \sigma^2_{\rm ADU} = \sigma^2_N + \sigma^2_{\rm read-out} $.
But since $\sigma^2_N = N$, we get 
\begin{equation}
    \frac{1}{g} \mathrm{ADU} + \left( \frac{\sigma^2_{\rm read-out}}{g^2} \right) = \sigma^2_{\rm ADU}
\end{equation}
Hence, we can compute $g$ and $\sigma_{\rm read-out}$ by measuring $\sigma^2_{\rm ADU}$ as a function of $\rm ADU$.
To do so, we took flat exposures at multiple exposure times within the linearity limit.
At each exposure time, we took two images, $\rm IM_{1,2}$. $\rm ADU(t)$ is the mean count of $\rm (IM_{1} + IM_{2})/2$ in the pair of images.
The associated variance ($\sigma^2_{\rm ADU(t)}$) is the count variance of $\rm (IM_{1} - IM_{2})/2$ in the image.
By measuring this at different exposure times, we could fit for $g$ and $\sigma_{\rm read-out}$, and arrived at $g = 1.2 \, \rm e^{-}/ADU$ and $\sigma_{\rm read-out} = 12\, \rm e^-$.

\subsection{Polarization grating orientation}\label{sec:orientation}
Recall that the QWP/PG device with the split-pupil design splits and disperses the incoming beam into 4 outgoing beams according to the incoming linear polarization states. 
To measure exactly what polarization angle each quadrant on the detector corresponds to, we observed the highly polarized twilight sky at zenith, where the polarization angle is perpendicular to the Sun's azimuth.
Aggregating multiple observations from different nights over the year, we found that the 0, 90, 45, and 135$^\circ$ linear polarization angle ($\rm Q_p$, $\rm Q_m$, $\rm U_p$, and $\rm U_m$) corresponds to the lower left, upper right, lower right, and upper left quadrants respectively. 
A more precise measurement of the angle of polarization is presented in \textsection\ref{sec:pol_star}.

\subsection{Instrument transmission}\label{sec:throughput}
\begin{figure*}
    \centering
    \includegraphics[width=0.49\textwidth]{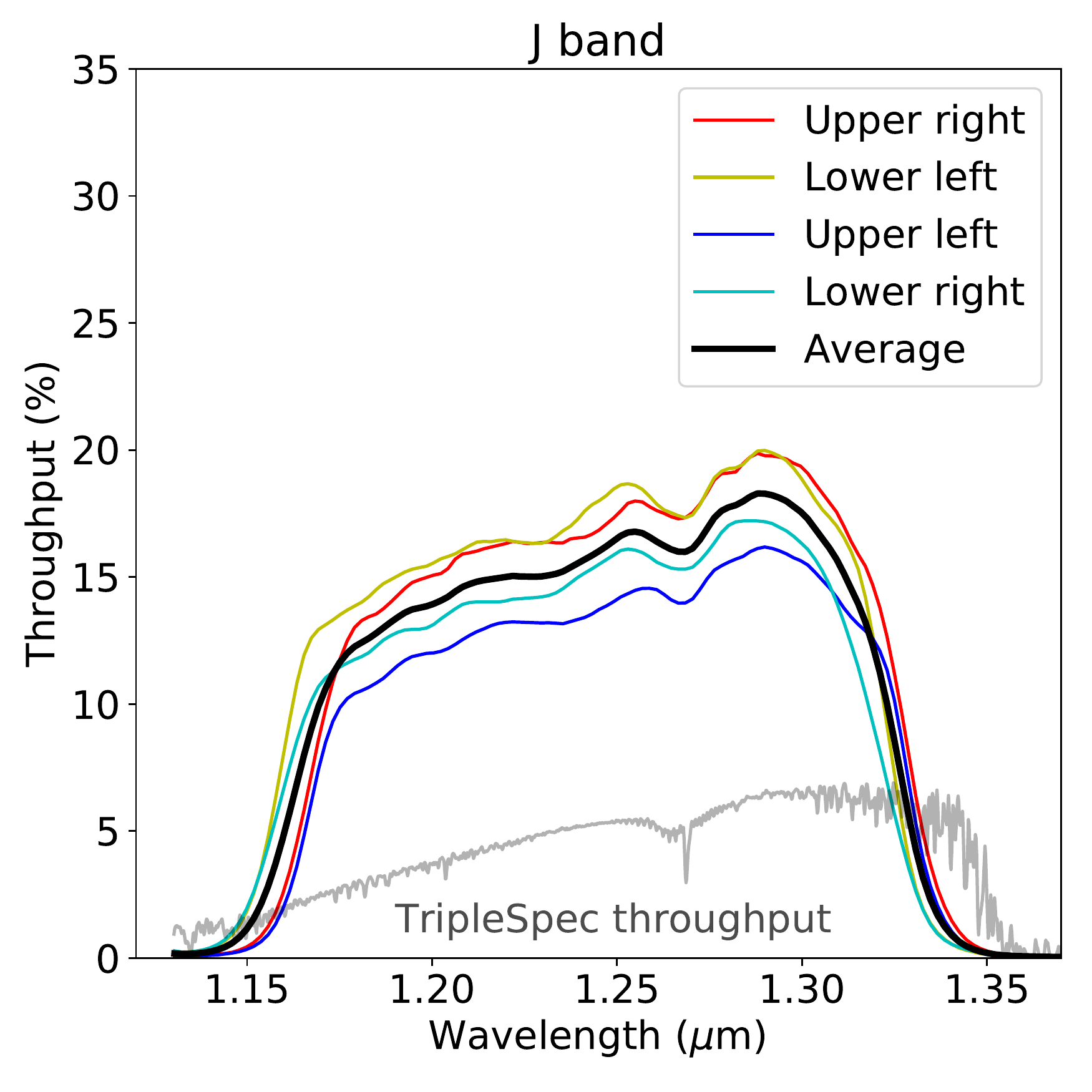}
    \hfill
    \includegraphics[width=0.49\textwidth]{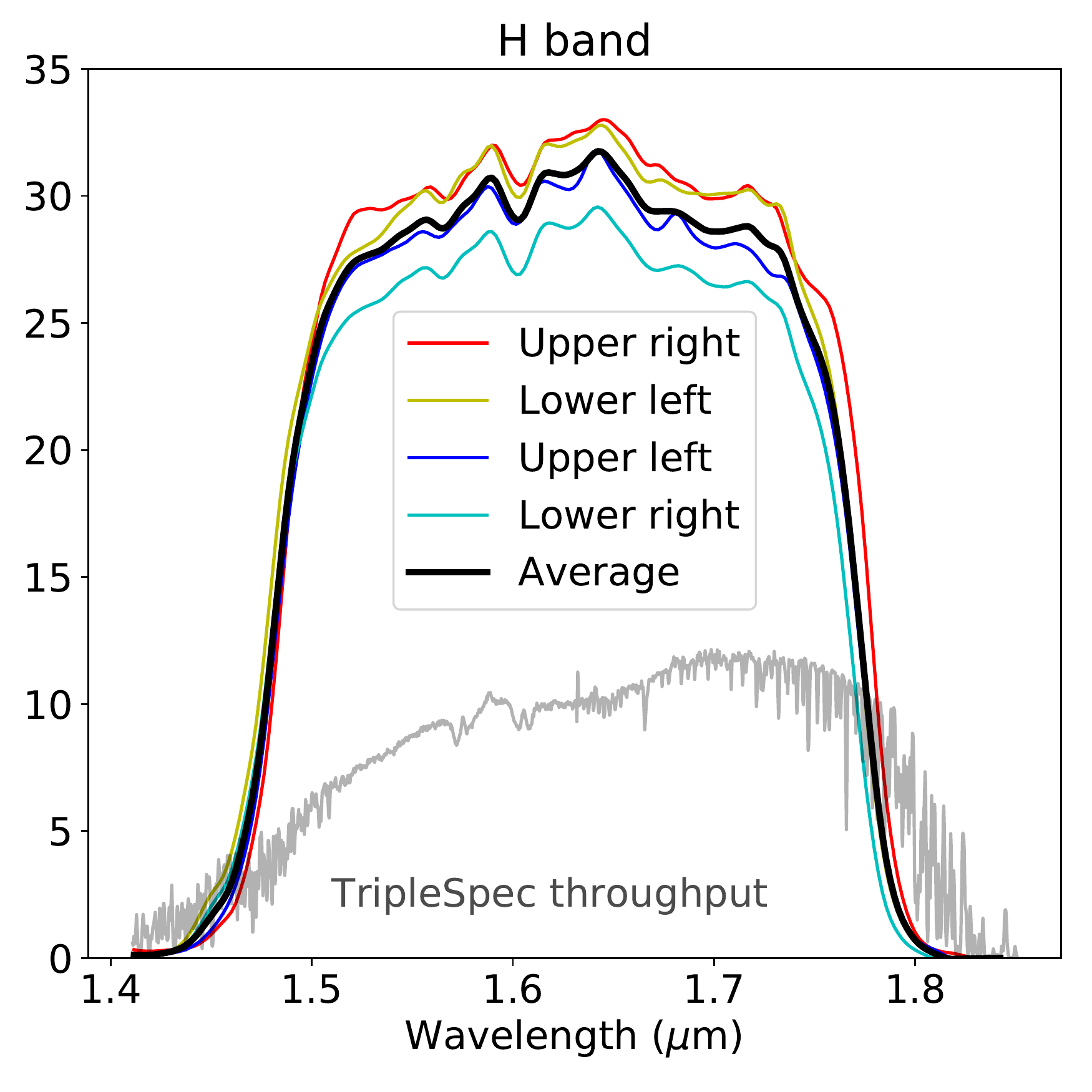}
    \caption{WIRC+Pol's transmission in the J band and the H band. Individual trace's transmission is computed from the ratio between 1/4 expected flux above the atmosphere to what is measured at the detector. The factor of 1/4 reflects the fact that we divide the incoming light into 4 beams for the 4 polarization angles. (Thus, if the transmission of the atmosphere and instrument were perfect, each trace would measure 100\% throughput in these plots). The average transmission, which corresponds to the total instrumental transmission from top of the atmosphere to the 4 spectral traces, is overplotted. TripleSpec's transmission is given for comparison, though TripleSpec has a higher spectral resolution and is much more optically complex. A few atmospheric absorption lines at 1.27, 1.57, and 1.61 $\mu$m visible in both TripleSpec and WIRC+Pol spectra in both J and H bands are used for confirming the wavelength solution. Other spectral features that are only present in WIRC+Pol's efficiency come from the broadband J and H filters. The relative shift of the filter transmission profiles for upper traces and lower traces is evident, especially for the J band, due to different angles of incidence on the broadband filter.}
    \label{fig:throughput}
\end{figure*}

We conducted two separate measurements in order to characterize both WIRC+Pol's absolute transmission from above the atmosphere to detector, and the transmission of just the PG.  
The absolute transmission can be measured by observing an unpolarized source for which we know the spectrum in physical units.
Comparing the spectrum observed by WIRC+Pol to this known spectrum allows us to measure the efficiency of photon transfer from top of the atmosphere to our detector. 
For this measurement, we first need a flux calibrated spectrum of an unpolarized source, observed and calibrated using a different instrument.
We observed unpolarized, A0 standard star HD\,14069 on 2017 October 12 using \emph{TripleSpec}, which is a medium resolution near-IR spectrograph at the Cassegrain focus of the 200-inch telescope that has simultaneous wavelength coverage from 0.9 to 2.4 $\mu$m, i.e. y, J, H, and K bands \citep{herter2008}. 
To flux calibrate the spectrum, we also observed an A0V standard star, HIP\,13917, at a similar airmass.
Raw spectra for both HD\,14069 and HIP\,13917 are reduced and extracted using a version of the Spextool data reduction pipeline, modified for Palomar TripleSpec \citep{cushing2004}. 
Finally, to remove telluric absorption and to flux calibrate the spectrum of HD\,14069, we use the \texttt{xtellcorr} tool \citep{vacca2003}, which derives TripleSpec's transmission by comparing the A0V model spectrum (derived from Vega) to the observed A0V spectrum.
This derived transmission, shown in Fig.~\ref{fig:throughput} for reference, is applied to HD\,14069's observed spectrum in the instrumental unit to get the spectrum in a physical flux unit. 

Next, we observed the same star using WIRC+Pol in the J band on 2017 October 16. 
The data were calibrated and extracted using the reduction pipeline described above, and we have four spectra in WIRC+Pol's instrumental unit ($\rm ADU\,s^{-1}$). 
Multiplying this spectrum by the gain and dividing by the width of each wavelength bin, we get the spectrum in $\rm e^-\, s^{-1}\, \mu m^{-1}$.
To get the TripleSpec spectra from the physical unit ($\rm erg\,s^{-1}\,cm^{-2}\,\mu m^{-1}$) into the same unit, we multiply it by the telescope collecting area and divide by the energy per photon. 
We then convolve this spectrum with a Gaussian kernel down to WIRC+Pol resolution. 
The ratio between these two spectra is the fraction of photons from this source from the top of the atmosphere reaching WIRC+Pol's detector. 
For the H band measurement, we observe a different star with the same spectral type (HD\,331891), and repeat the analysis with the TripleSpec spectrum scaled for the new source.

Fig.~\ref{fig:throughput} shows the transmission of each of the four WIRC+Pol spectral traces (note that the total flux is divided into 4 traces). 
The average transmission is overplotted. 
TripleSpec's transmission, measured by our observations described above, is given for reference. 
The number is about a factor of 2 lower than previous measurements by \citet{herter2008}, which may be due to the different atmospheric conditions. 
We note that WIRC+Pol has a very high transmission, peaking at 17.5\% and 30\% in J and H bands respectively. 
The four spectral traces have different relative transmission, which mimics an effect of instrumental polarization. 
We will discuss this issue in the next section, but this effect necessitates observations of an unpolarized standard star.
The $\rm O_2$ and $\rm CO_2$ atmospheric absorption features in the J and H bands that we used to align the four spectral traces, as mentioned in \textsection\ref{sec:pol_computation}, are visible in both WIRC+Pol's and TripleSpec's transmission curves. 
Additional features in WIRC+Pol's transmittance curve are due to the broadband filters.
Finally, we note that TripleSpec's transmission has a strong wavelength dependence, intrinsic to a surface relief grating, while WIRC+Pol's transmission is almost flat.
(The J band slope is due to the telescope mirror coating, see Fig. 2 in \citet{herter2008}.)

In addition to the absolute transmission of the instrument, we also measured the transmission of the PG itself by observing a bright star (HD\,43384) with and without the PG.
We dark subtract and flat divide the raw data, then median combine images with and without the PG.
We performed aperture photometry using an Astropy \citep{astropy2018} affiliate \texttt{photutils} package to compare flux in the direct image without PG to flux in the spectral traces with PG. 
In an ideal scenario, all four traces will get an equal amount of flux, which is the direct flux divided by four.
However, the measurement shows that the $\rm Q_p$ (lower left), $\rm Q_m$ (upper right), $\rm U_p$ (lower right), and $\rm U_m$ (upper left) have the efficiency of 88.3, 84.4, 98.7, and 99.2\%, in comparison to the ideal scenario. 
Note that these numbers are consistent to what we found in the absolute transmission measurement.
The difference between the $Q$ and $U$ pair transmission is likely due to the misalignment between the pupil plane and the WIRC instrument. 
This misalignment is also responsible for $\sim$20\% gradient in the flat field taken without the PG. 
We then assume that this difference is not due to an intrinsic difference between the transmission of the two halves of the PG.
Thus we report its mean transmission as 93\%.

%

\subsection{Observations of unpolarized standard stars}\label{sec:unpol}
\begin{figure*}
    \centering
    \includegraphics[width=0.6\textwidth]{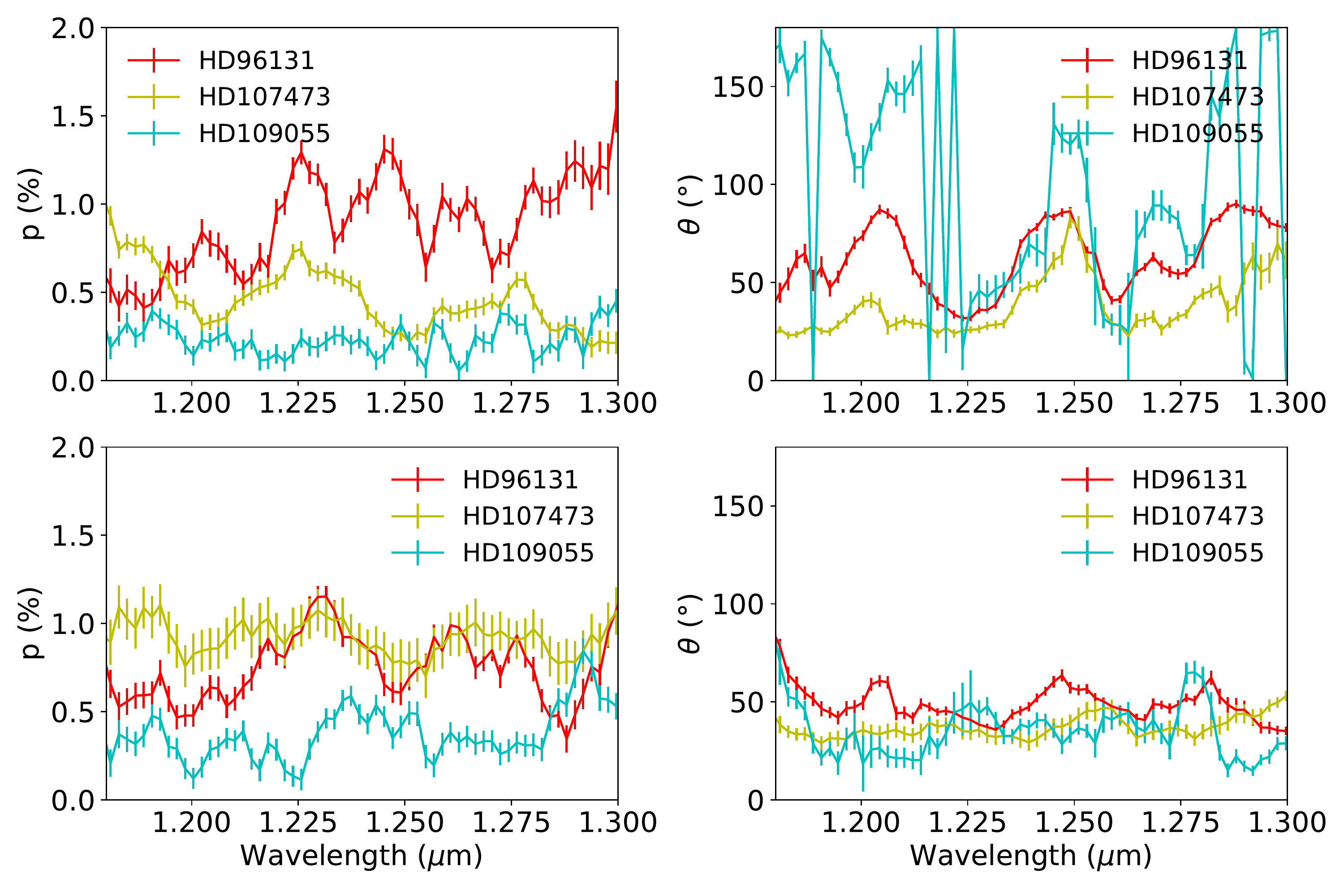} 
    \includegraphics[width=0.39\textwidth]{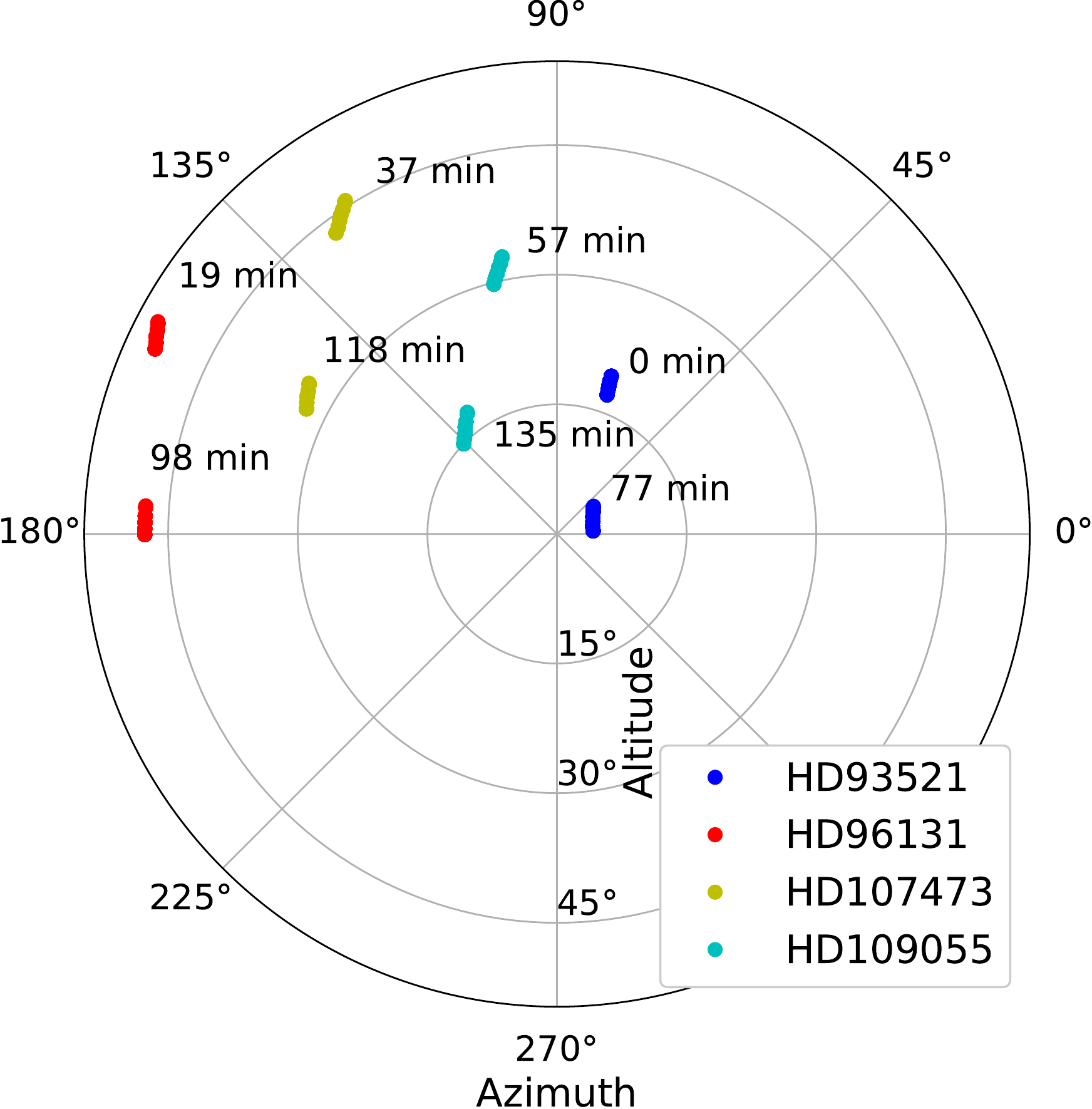}
    \caption{\textit{Left} four panels show measured degree and angle of polarization of 3 unpolarized standard stars HD\,96131, HD\,107473, and HD\,109055 using HD\,93521 as the standard (see text for reduction details). The top and bottom rows are from the first and second sets of observations. 
    \textit{Right} The location of the stars on sky during the observing sequence along with the time between the beginning of each sequence and the beginning of the first sequence is annotated. 
    The black dots represent the 3-hour long sequence observation of HD\,109055 on 2018 May 04.
    Lines of constant hour angle are plotted.}
    \label{fig:four_unpol}
\end{figure*}

In order to quantify the instrumental polarization due to telescope pointing, we observed 4 different unpolarized standard stars: HD\,93521, HD\,96131, HD\,107473, and HD\,109055 \citep{heiles2000} on 2018 April 21.
All stars are polarized to less than 0.1\% in the V band, which yield negligible polarization in the IR assuming Serkowski law, $p(\lambda)/p(\lambda_{\rm max}) = \exp{(-1.15 \ln^2{(\lambda/\lambda_{\rm max})})}$\citep{serkowski1975}. 
We observed the four stars in the aforementioned order, then repeat the observations in the same order so each star was visited twice. 
Fig.~\ref{fig:four_unpol} \textit{right} shows the location of these 4 stars on the sky in altitude-azimuth coordinates (which reflect gravity vector on the instrument).
Hour angles in 2 hour interval are plotted as well. 
The total exposure time per visit is 500-600 s, resulting in typical SNR for the spectra of order 3,000 for HD\,93521 and HD\,107474 ($J\sim 7.5$) and 1,500 for HD\,96131 and HD\,109055 ($J\sim 8.8$). 
For each of the two visits to the stars, we used HD\,93521 as the ``standard'' ($S'_{p,m}$ in \eqref{eq:qu}) and the remaining 3 stars as the ``source'' ($I'_{p,m}$ in \eqref{eq:qu}).
The resulting measured $q$ and $u$ are the difference between instrumental polarization between the two standard stars. 
We then used HD\,93521 observations from the two visits to calibrate each other. This provides us the first handle of the temporal stability of the instrumental polarization, which shall be discussed in greater details in \textsection\ref{sec:stability}. 
Fig.~\ref{fig:four_unpol} \textit{left} four panels show the measured degree and angle of polarization measured from these observations while the \textit{right} panel shows the locations of the 4 stars on sky in the two sequences.
The time delay between the first observation in each sequence and the beginning of the observation is annotated. 
Out of the 3 stars compared with HD\,93521, only HD\,109055 results in measured polarization consistent with zero to within 3$\sigma$. 
The other two stars show deviation up to 1\%. 
We note that for both sequences, HD\,96131 and HD\,107473 were observed \textit{closer} in time to HD\,93521, however, they were \textit{further} away on sky.
The intrinsic spectral type and brightness difference between these sources should not influence our reduction using the methods outlined above. 
Indeed, the deviation of measured polarization from zero did not seem to be a function of source's intrinsic properties.
HD\,93521 and HD\,109055, the pair that provided near-zero polarization differ in magnitude ($J = 7.5$ vs 8.7) and spectral type (O9.5IIInn vs A0V). 
This preliminary work led us to conclude that on sky pointing may have a noticeable effect on the measured polarization, and has implications for our future observation strategy: to observe the unpolarized standard star closest to the source. 
This may be results of differential atmospheric effects from observations at different airmass, or stress induced birefringence from the changing gravity vector on the instrument at different telescope pointing.

\subsection{Polarization spatial stability}\label{sec:ip}

As discussed earlier, we expect the polarization measurement of an unpolarized source to be non-zero due to instrumental systematics.
This may be due to an intrinsic telescope or instrument induced polarization or simply uncorrected flat field variation.
To quantify this effect, we mapped the polarization variability across the field of view by observing an A0 unpolarized standard star (HD\,14069) in a grid across the full field of view on 2017 Nov 28.
However, the observations were taken at a relatively low SNR and over a long period of time where other factors may affect the measured polarization.
While the fidelity of the measurements was not enough to construct a precise model of the polarization zero point as a function of location on the field of view, we found enough evidence that the polarization zero point can vary more than 1\% across the field of view.
This finding informed our decision to observe sources at one specific location on the detector to reduce this effect. 
(Each quadrant of the detector is split into four triangular regions by the focal plane mask (see Fig. \ref{fig:raw_data}), we pick the bottom triangle because of the general lack of bad pixels there.)

\begin{figure*}
    \centering
    \includegraphics[width = \textwidth]{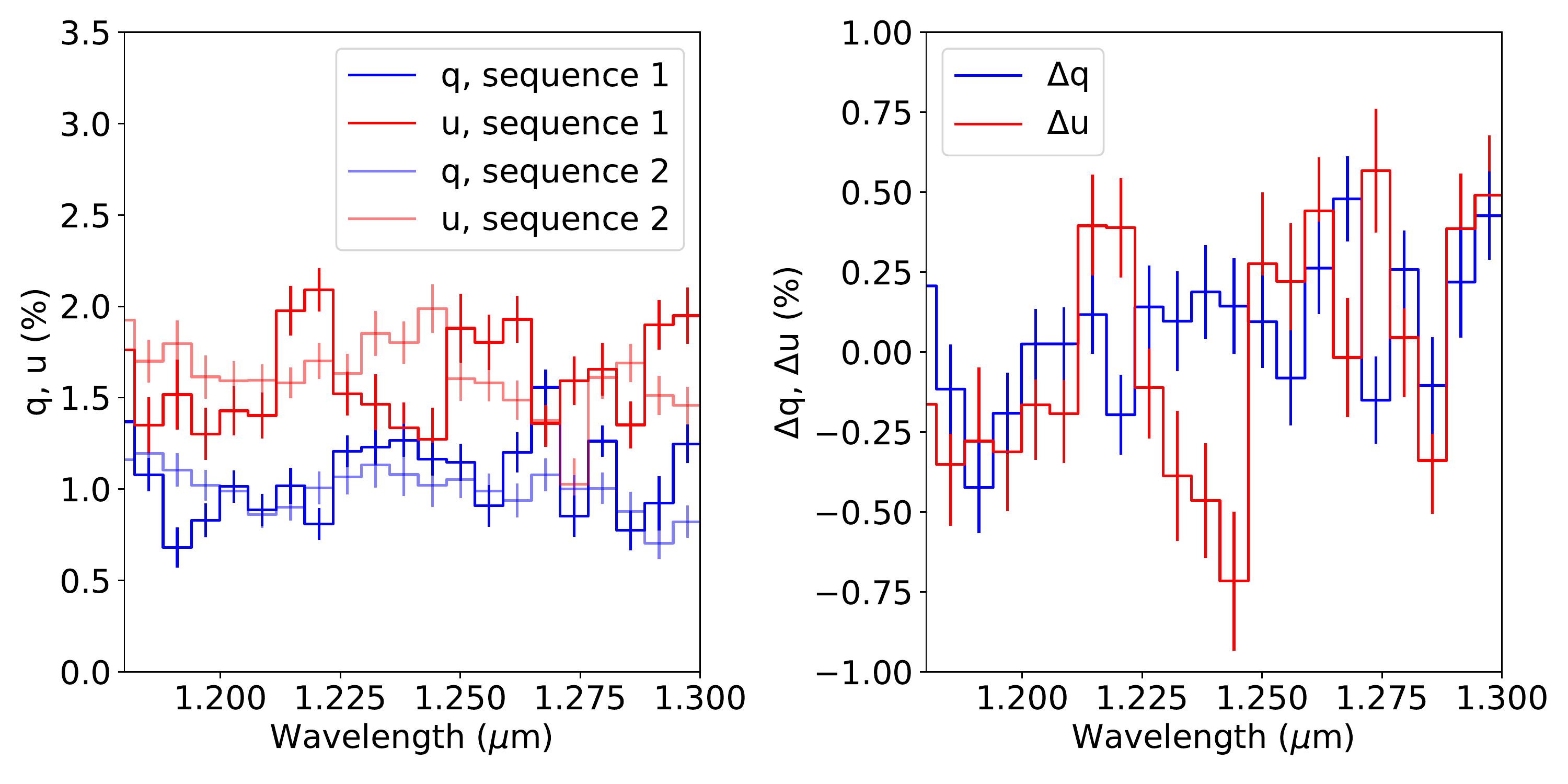}
    \caption{\textit{Left} $q$ and $u$ measured by using observations at position A to calibrate observations at position B. 
    Solid lines are the results from the first sequence of observations while the transparent lines are the second sequence.
    These measured $q$ and $u$ reflect the difference between instrumental effects at the two positions.
    \textit{Right} The difference between measured $q$ and $u$ from the two sequences.}
    \label{fig:AB}
\end{figure*}

In order to better quantify the spatial dependence of the instrumental effect, on 2018 July 24 we observed an F8V unpolarized standard star HD\,154892 at two dither positions on the detector (``A'' and ``B'').
We first took a sequence of 18 exposures, 100 s each, switching between A and B positions with an offset of 25" after every image. 
Three hours later, we conducted a similar observation of HD\,154892 at the same location on the detector with 20 exposures, 100 s each, switching between A and B positions.
In this sequence, the A position is the same as the A position in sequence 1, however, the offset size was 30". 
For each sequence of the observations, we median combine all spectra from positions A and B separately. 
Then we use position A as the standard (i.e. $S'_{p,m}$ in \eqref{eq:qu}) to calibrate observations from position B (i.e. $I'_{p,m}$ in \eqref{eq:qu}).
The measured $q$ and $u$ are then the \textit{difference} between instrumental $q$ and $u$ at positions A and B. 
Fig.~\ref{fig:AB} \textit{left} shows $q$ and $u$ differences between positions A and B for sequence 1 (solid line) and 2 (transparent line). 
The difference in instrumental polarization between these two positions are 1.0\% and 1.5\% in $q$ and $u$ respectively.  
Fig.~\ref{fig:AB} \textit{right} shows the difference between the two sequences, which quantify the temporal stability of the spatial systematic difference. 
While the average over the J band of the difference is around 0, some wavelength dependence exists.
This may be from the fact that the offset between positions A and B was slightly difference between sequence 1 and 2 (25" vs 30"), or it could be a real temporal change in instrumental polarization spatial dependence. 
In summary, this measurement shows that the spatial dependence of the instrumental polarization is of order 1-1.5\% over 30" on the detector, and this difference is temporally stable to $\pm\sim$0.3\%. 
This finding underlines the need to observe a standard star and a science source at the exact same position on the detector, which can be done using the guiding script discussed in \textsection\ref{sec:upgrade}.

\subsection{Polarization temporal stability}\label{sec:stability}
In this section we quantify the temporal stability of the systematic polarization.
For instance, if we know that the 1\% instrumental polarization can be well measured and is stable at 0.1\% level over some period of time, then we can use observations of unpolarized standard stars to remove this systematic error and recover the source's true polarization down to $\sim$ 0.1\% level.
Hence, we need to quantify the timescale over which our instrumental polarization zero point changes.

\begin{figure*}
	\includegraphics[width = \linewidth]{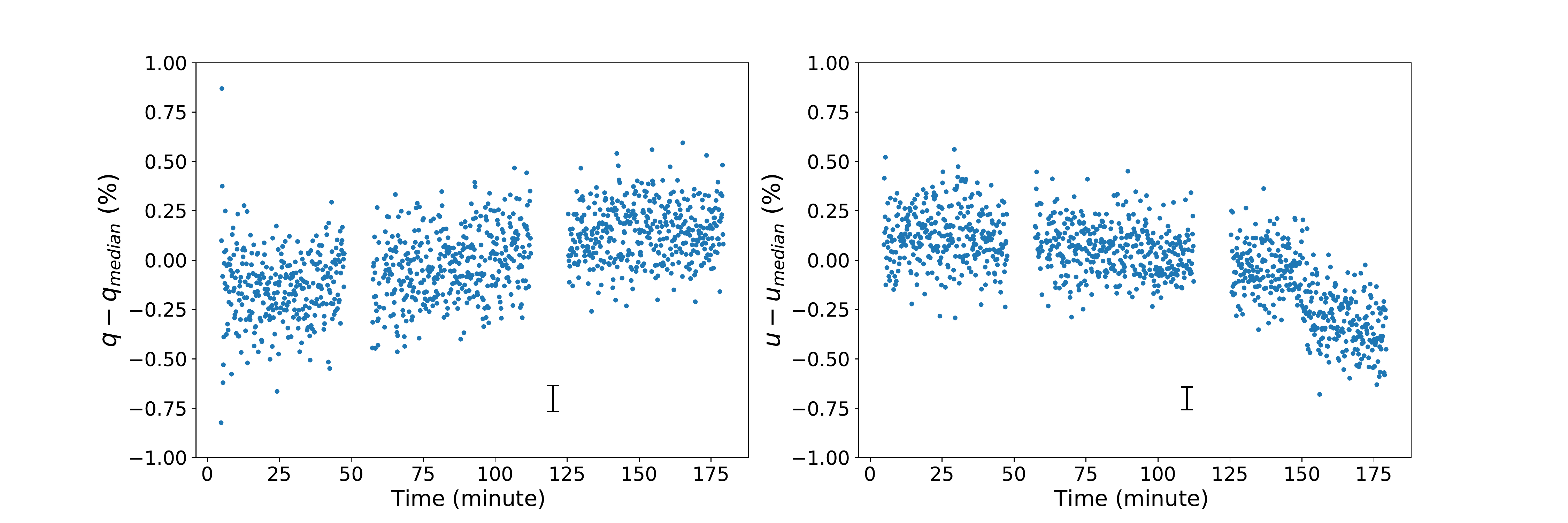}
    \caption{Temporal variation of the measured broadband normalized Stokes parameters $q$ (\textit{left}) and $u$ (\textit{right}) as functions of time since the first image. The broadband value is simply the median of $q$ and $u$ spectra within the J band bandpass. Each data point plotted here comes from an individual image taken in the sequence, and is subtracted by the median broadband $q$ and $u$ across the whole observing sequence. Uncertainty in $q$ and $u$ remains constant, and a 1$\sigma$ representative errorbar is shown in each plot. The two gaps in the data at 50 and 120 minutes are when we refocused the telescope. 
    }
    \label{fig:temp_stability}
\end{figure*}

To conduct this measurement, we observed an A0V unpolarized (0.07$\pm$0.07\% in the V band, consistent to zero) standard star (HD\,109055; \citealp{heiles2000}) on 2018 May 04 UT for 3 hours as the star traces 45$^\circ$ of telescope pointing angle in hour angle across the meridian. 
The on-sky location of HD\,109055 is shown as black dots in Fig.~\ref{fig:four_unpol}.
Our guiding script kept the source on a single point on the detector with guiding RMS $\sim$ 0.25" (1 pixel) to reduce the field of view dependent effects.  
We refocused the telescope twice during the observing sequence to keep up with the changing temperature inside the dome since our observations happened at the beginning of the night, which show up as gaps in our time series in Fig. \ref{fig:temp_stability}.
The data were reduced by the DRP using the procedure described above (\textsection\ref{sec:drp}).
We first median combined all the spectra of the source, from which we computed median $q_{\rm median}$ and $u_{\rm median}$ to provide a baseline. 
Next we compute $q_i$ and $u_i$ spectra from each of the single observations, and $q_i - q_{\rm median}$,  $u_i - u_{\rm median}$ shows the variation in the polarimetric zero point throughout our 3-hour long observing sequence. 
We found that the seeing conditions remain very stable and the polarimetric deviation in both $q$ and $u$ show no wavelength dependence, which may happen if the spectral resolution of the trace is changing due to seeing variations.
Hence, for each observation, we use the median of $q_i - q_{\rm median}$ and $u_i - u_{\rm median}$ within the filter's bandpass as broadband values, shown in Fig. \ref{fig:temp_stability}.
The two gaps in the data indicate where we refocused the telescope.
The RMS of the variation is 0.2\% for both $q$ and $u$ over 30 minutes, corresponding to 0.13 range in airmass. 
Note that there are some long term variations, whose origin remain uncertain.  
We note that the change in systematic polarization due to telescope pointing (discussed in \textsection\ref{sec:unpol}) is quantitatively consistent with what we observed in this long sequence.
While the telescope pointing effect contributes to the long term variation in the systematic presented here, there might also be other components that are still unknown.


\subsection{Observations of known polarized stars}\label{sec:pol_star}
\begin{figure*}
    \centering
    \includegraphics[width = 0.8\linewidth]{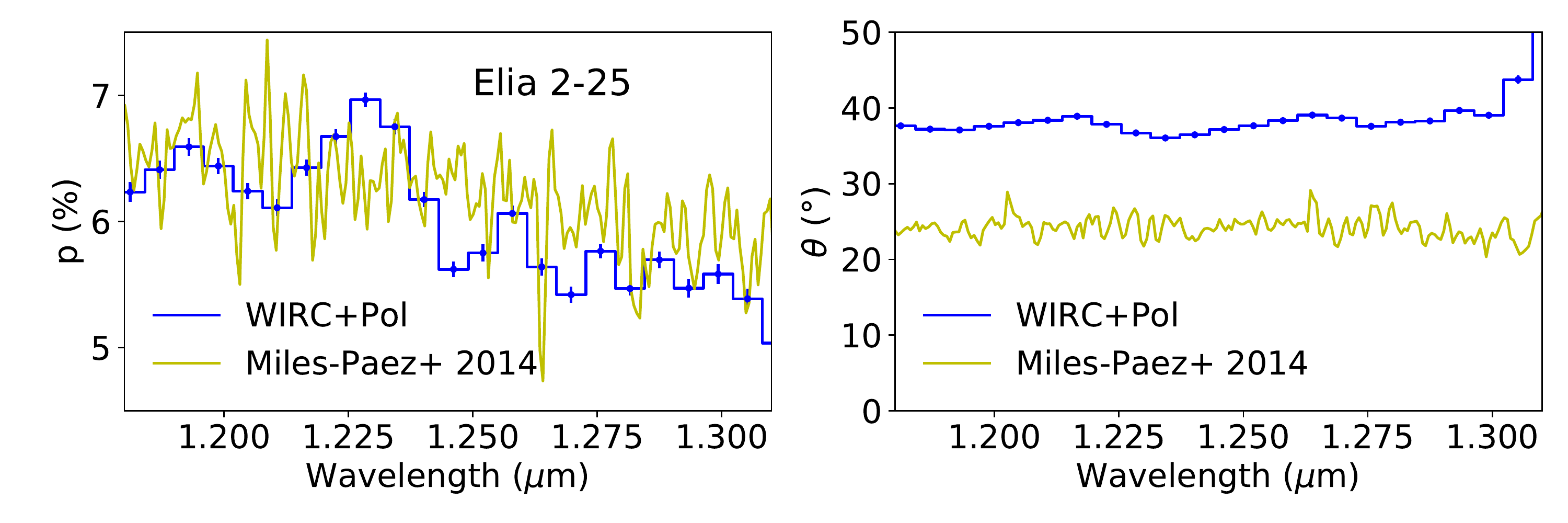}
    \includegraphics[width = 0.8\linewidth]{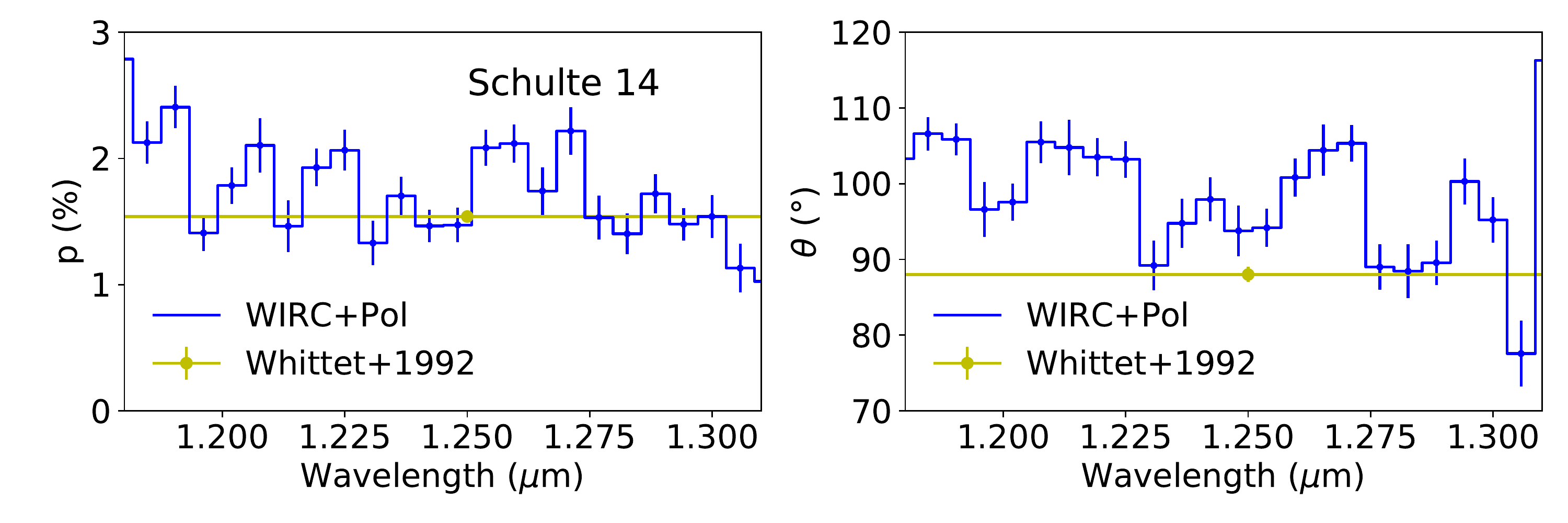}
    \caption{Degree ($p$) and angle ($\theta$) of polarization for Elia\,2-25 (\textit{top}) and Schulte\,14 (\textit{bottom}) from WIRC+Pol in comparison to the results from \cite{miles2014} or \cite{whittet1992}. 
    The y-axes range for the two stars are the same.
    The degree of polarization agrees to within 0.5\%, but the angle of polarization is $\sim$15$^\circ$ off. 
    Error bars only account for random errors, which appear to be smaller than typical scattering in $p$ and $\theta$ values. 
    This is likely due to systematic error in aligning spectra to compute polarization.}
    \label{fig:elia2-25}
\end{figure*}

Once the polarimetric zero point is well characterized, observations of stars with known polarization are required to measure the instrument's polarimetric efficiency and polarization angle zeropoint. 
The first star used was Elia\,2-25, which is a polarized standard in \citet{whittet1992} with $p = 6.46 \pm 0.02\%$ and $\theta = 24 \pm 1^\circ$ in the J band.
It has near-IR polarization spectrum published by \citet{miles2014}.  
We observed Elia 2-25 \citep{miles2014, whittet1992} on 2018 May 06 for 17 min (10 min), followed immediately by an unpolarized standard HD\,154892 \citep{heiles2000} for 8 min (2 min), both wall clock time (total integrated time). 
Both stars were put to within a pixel from each other on the detector to minimize the spatially dependent polarization effect discussed above. 
The total time of 25 min is short enough for the calibration to not be affected by the varying systematic shown in the previous section. 
Fig.~\ref{fig:elia2-25} shows the degree of polarization ($p$ in percent) and the angle of polarization ($\theta$ in degrees), in comparison from the literature result.
The degree of polarization agrees to the literature value to within 0.5\% across the whole spectrum, but the angle of polarization is greater than the literature value by 15$^\circ$. 
We know that the instrument is aligned with North up to within 1$^\circ$ by observations of star trails, so this offset must be from the instrument itself. 
The second polarized standard observed was Schlute\,14 with $p = 1.54 \pm 0.02\%$ and $\theta = 88 \pm 1^\circ$ in the J band \citep{whittet1992}.
Fig.~\ref{fig:elia2-25} \textit{bottom} shows the measured polarization compared to the literature. 
The results agree to those from Elia\,2-25, with $p$ accurate to within 0.5\%. 
We note that the agreement between WIRC+Pol observations and literature values to within 0.5\% is consistent to the systematic polarization due to telescope pointing as discussed in \textsection\ref{sec:unpol}, since the unpolarized standards used here were not spatially close to the polarized standards on sky. 
The angle of polarization, however, is offset from the literature value by 15\textdegree.
We know from observing star trails on WIRC, with the telescope tracking off, that the orientation of instrument, since the PG/QWP, is offset from the North (0\textdegree angle of polarization) by $\sim$1\textdegree.
The most likely culprit of the offset is the angle of polarization zero point intrinsic to the PG/QWP device.
In another word, the PG/QWP device was manufactured to sample 15\textdegree, 60\textdegree, 105\textdegree, and 150\textdegree instead of the anticipated 0\textdegree, 45\textdegree, 90\textdegree, and 135\textdegree.
As a result, it simply rotates the angle of polarization measurement by 15\textdegree and did not affect the degree of polarization.
As a result, we can measure and subtract this offset during the course of an observation.

\section{Future instrument upgrades}\label{sec:future_upgrade}

Informed by these commissioning results, we have identified a few potential upgrades that would improve the instrument's performance. 
\begin{itemize}
    \item An addition of a polarimetric modulator---a device that can rotate the incoming beam's polarization angle---will allow us to measure linear polarization from each of the four spectral traces using four different modulation angles. 
    The modulator will allow us to swap the incident polarization between different pairs of spectral traces, while the instrumental systematics stay constant. Thus allowing us to distinguish between true astrophysical polarization and instrumental systematics. 
    This upgrade would remove the observed field dependent polarimetric zero point and other slowly varying effects (\textsection\ref{sec:stability}, Fig. \ref{fig:temp_stability}).
    The upgrade has been funded and will be implemented by the end of 2018.
	\item To minimize non-common path errors between the four spectral traces, the PG has to be the last optic in the optical train before the reimaging optics. 
    This can be done by swapping the PG and broadband filters, which is a complicated process since the two filter wheels are not interchangeable, and the PG requires a special mounting on the filter wheel.
    Another solution to this problem is to place a J+H band filter permanently in front of the PG.
    This way, the instrument will be able to observe in the J and H bands simultaneously with the caveat of a brighter sky background in the slitless mode.
    We note that this change may not be needed with the presence of a modulator. 
    
\end{itemize}

\section{Conclusion}\label{sec:conclusion}

We described a R$\sim$100 near-IR spectropolarimeter, WIRC+Pol, on the 200-inch telescope at Palomar Observatory.
The existing IR imaging camera, WIRC, was upgraded by an installation of a compact, liquid crystal polymer-based polarimetric device called a PG.
The PG acts both as a polarimetric beam-splitter and a spectral disperser, and is small enough to fit inside the filter wheel of the instrument, simplifying the upgrade in comparison to using a Wollaston prism and another grating.
We developed a data reduction pipeline that extracts spectra from the images and computes polarization of the observed source.

We have established the following key characteristics of the instrument. 
Firstly, the liquid-crystal based QWP/PG device performs as expected, delivering a high dispersion efficiency of 93\% into the first order spectra.  
This is an on-sky demonstration that a PG, apart from its polarimetric capabilities, is a very efficient disperser in comparison to a surface relief grating.
Secondly, the commissioning data showed that the instrument can measure linear polarization reliably to 1\% level for bright sources with known polarization given an appropriate observation strategy.
The measured polarization angle is greater than literature values by $\sim 15^\circ$, which is constant and can be removed. 
The polarimetric uncertainty is currently limited by time-varying systematics, which may originate from telescope pointing, likely due to stress induced instrumental polarization or atmospheric effects. 
Thirdly, we documented difficulties of computing polarization from single-shot observations without a rotating modulator.
Relying on comparing fluxes in four spectral traces in four quadrants of the detector to compute polarization risks confusing source's intrinsic polarization with instrument's flat field and non-common path errors.
We mitigated this effect by correcting our observations with deep flat field images taken with all polarimetric optics in place, and also by keeping the source in all observations on a single location on the detector to within 1 pixel (0.25"). 
Another requirement to compute polarization from comparing fluxes in four spectra is that they must be well aligned in the wavelength direction.
This was complicated by the fact that the broadband filters used are downstream from the PG, imprinting different transmission profiles on the four traces. 
This was mitigated by using atmospheric absorption features to align the spectra instead of using the filter cutoffs. 
The presented characterization of WIRC+Pol was crucial to inform the funded half-wave plate instrument upgrade in the near future. 
The discovered characteristics should inform the design of a future spectropolarimetric instrument using a PG. 
The lack of a rotating modulator in WIRC+Pol may have caused a number of systematics, but this design can provide a very efficient spectropolarimeter with minimal moving parts, which may prove essential in incorporating such system in a future space-based instrument.

\acknowledgments 
This work is supported by the National Science Foundation under Grant No. AAG-1816341. 
The WIRC+Pol upgrade was in part supported by a grant from the Mt. Cuba Astronomical Foundation.
We thank Paulo Miles-P\'aez for providing us polarized spectra of Elia 2-25 polarized standard star for comparison with our observations and for discussions on best practices in reducing near-IR spectropolarimetric data.
We thank staff of Palomar Observatory for assisting our observations. 
Palomar Observatory is operated by a collaboration between California Institute of Technology, Jet Propulsion Laboratory, Yale University, and National Astronomical Observatories of China. 

\bibliography{main} 

\newcommand{\noop}[1]{}
\begin{thebibliography}{}
\expandafter\ifx\csname natexlab\endcsname\relax\def\natexlab#1{#1}\fi
\providecommand{\url}[1]{\href{#1}{#1}}

\bibitem[{{Bertin} \& {Arnouts}(1996)}]{Bertin1996}
{Bertin}, E., \& {Arnouts}, S. 1996, \aaps, 117, 393

\bibitem[{{Cushing} {et~al.}(2004){Cushing}, {Vacca}, \&
  {Rayner}}]{cushing2004}
{Cushing}, M.~C., {Vacca}, W.~D., \& {Rayner}, J.~T. 2004, \pasp, 116, 362

\bibitem[{{de Kok} {et~al.}(2011){de Kok}, {Stam}, \& {Karalidi}}]{dekok2011}
{de Kok}, R.~J., {Stam}, D.~M., \& {Karalidi}, T. 2011, \apj, 741, 59

\bibitem[{{Debes} {et~al.}(2016){Debes}, {Ygouf}, {Choquet}, {Hines}, {Perrin},
  {Golimowski}, {Lajoie}, {Mazoyer}, {Pueyo}, {Soummer}, \& {van der
  Marel}}]{debes2016}
{Debes}, J.~H., {Ygouf}, M., {Choquet}, E., {et~al.} 2016, Journal of
  Astronomical Telescopes, Instruments, and Systems, 2, 011010

\bibitem[{{Escuti} {et~al.}(2006){Escuti}, {Oh}, {S\'anchez}, {Bastiaansen}, \&
  J.}]{escuti2006}
{Escuti}, M.~J., {Oh}, C., {S\'anchez}, C., {Bastiaansen}, C., \& J., B.~D.
  2006, in , 6302 -- 6302 -- 11.
\newblock \url{https://doi.org/10.1117/12.681447}

\bibitem[{{Ghinassi} {et~al.}(2002){Ghinassi}, {Licandro}, {Oliva}, {Baffa},
  {Checcucci}, {Comoretto}, {Gennari}, \& {Marcucci}}]{ghinassi2002}
{Ghinassi}, F., {Licandro}, J., {Oliva}, E., {et~al.} 2002, \aap, 386, 1157

\bibitem[{{Heiles}(2000)}]{heiles2000}
{Heiles}, C. 2000, \aj, 119, 923

\bibitem[{{Herter} {et~al.}(2008){Herter}, {Henderson}, {Wilson}, {Matthews},
  {Rahmer}, {Bonati}, {Muirhead}, {Adams}, {Lloyd}, {Skrutskie}, {Moon},
  {Parshley}, {Nelson}, {Martinache}, \& {Gull}}]{herter2008}
{Herter}, T.~L., {Henderson}, C.~P., {Wilson}, J.~C., {et~al.} 2008, in
  \procspie, Vol. 7014, Ground-based and Airborne Instrumentation for Astronomy
  II, 70140X

\bibitem[{{Horne}(1986)}]{Horne1986}
{Horne}, K. 1986, \pasp, 98, 609

\bibitem[{{Jensen-Clem} {et~al.}(2016){Jensen-Clem}, {Millar-Blanchaer},
  {Mawet}, {Graham}, {Wallace}, {Macintosh}, {Hinkley}, {Wiktorowicz},
  {Perrin}, {Marley}, {Fitzgerald}, {Oppenheimer}, {Ammons}, {Rantakyr{\"o}},
  \& {Marchis}}]{jensen2016}
{Jensen-Clem}, R., {Millar-Blanchaer}, M., {Mawet}, D., {et~al.} 2016, \apj,
  820, 111

\bibitem[{Keller(2001)}]{keller2001}
Keller, C.~U. 2001, in Astrophysical Spectropolarimetry:, ed.
  J.~Trujillo-Bueno, F.~Moreno-Insertis, \& F.~Sanchez (Cambridge: Cambridge
  University Press), 303--354

\bibitem[{Kirkpatrick(2005)}]{Kirkpatrick2005}
Kirkpatrick, J.~D. 2005, Annu. Rev. Astron. Astrophys, 43, 195.
\newblock
  \url{http://www.annualreviews.org/doi/pdf/10.1146/annurev.astro.42.053102.134017}

\bibitem[{{Larkin} {et~al.}(2003){Larkin}, {Quirrenbach}, {Krabbe}, {Aliado},
  {Barczys}, {Brims}, {Canfield}, {Gasaway}, {LaFreniere}, {Magnone},
  {Skulason}, {Spencer}, {Sprayberry}, \& {Weiss}}]{larkin2003}
{Larkin}, J.~E., {Quirrenbach}, A., {Krabbe}, A., {et~al.} 2003, in \procspie,
  Vol. 4841, Instrument Design and Performance for Optical/Infrared
  Ground-based Telescopes, ed. M.~{Iye} \& A.~F.~M. {Moorwood}, 1600--1610

\bibitem[{{Manchado} {et~al.}(2004){Manchado}, {Barreto}, {Acosta-Pulido},
  {Ballesteros}, {Barreto}, {Cadavid}, {Carrillo}, {Charcos}, {Correa},
  {Delgado}, {Dominguez-Tagle}, {Gonzalez}, {Hernandez}, {Lopez}, {Moreno},
  {Olives}, {Peraza}, {Prada}, {Redondo}, {Sanchez}, {Sosa}, {Tenegi}, \&
  {Vidal}}]{manchado2004}
{Manchado}, A., {Barreto}, M., {Acosta-Pulido}, J., {et~al.} 2004, in
  \procspie, Vol. 5492, Ground-based Instrumentation for Astronomy, ed.
  A.~F.~M. {Moorwood} \& M.~{Iye}, 1094--1104

\bibitem[{{Marley} \& {Sengupta}(2011)}]{marley2011}
{Marley}, M.~S., \& {Sengupta}, S. 2011, \mnras, 417, 2874

\bibitem[{{Miles-P{\'a}ez} {et~al.}(2014){Miles-P{\'a}ez}, {Pall{\'e}}, \&
  {Zapatero Osorio}}]{miles2014}
{Miles-P{\'a}ez}, P.~A., {Pall{\'e}}, E., \& {Zapatero Osorio}, M.~R. 2014,
  \aap, 562, L5

\bibitem[{{Millar-Blanchaer} {et~al.}(2014){Millar-Blanchaer}, {Moon},
  {Graham}, \& {Escuti}}]{millar2014}
{Millar-Blanchaer}, M., {Moon}, D.-S., {Graham}, J.~R., \& {Escuti}, M. 2014,
  in \procspie, Vol. 9151, Advances in Optical and Mechanical Technologies for
  Telescopes and Instrumentation, 91514I

\bibitem[{{Nagao} {et~al.}(2017){Nagao}, {Maeda}, \& {Tanaka}}]{nagao2017}
{Nagao}, T., {Maeda}, K., \& {Tanaka}, M. 2017, \apj, 847, 111

\bibitem[{{Nagao} {et~al.}(2018){Nagao}, {Maeda}, \& {Tanaka}}]{nagao2018}
---. 2018, \apj, 861, 1

\bibitem[{{Oliva}(1997)}]{oliva1997}
{Oliva}, E. 1997, \aaps, 123, 589

\bibitem[{{Packham} {et~al.}(2010){Packham}, {Escuti}, {Ginn}, {Oh}, {Quijano},
  \& {Boreman}}]{packham2010}
{Packham}, C., {Escuti}, M., {Ginn}, J., {et~al.} 2010, \pasp, 122, 1471

\bibitem[{{Patat} \& {Romaniello}(2006)}]{patat2006}
{Patat}, F., \& {Romaniello}, M. 2006, \pasp, 118, 146

\bibitem[{{Sengupta} \& {Marley}(2009)}]{sengupta2009}
{Sengupta}, S., \& {Marley}, M.~S. 2009, \apj, 707, 716

\bibitem[{{Sengupta} \& {Marley}(2010)}]{sengupta2010}
---. 2010, \apjl, 722, L142

\bibitem[{Serabyn {et~al.}(2016)Serabyn, Liewer, \& Mawet}]{serabyn2016}
Serabyn, E., Liewer, K., \& Mawet, D. 2016, Optics Communications, 379, 64.
\newblock
  \url{https://www.sciencedirect.com/science/article/pii/S0030401816304047?via{\%}3Dihub}

\bibitem[{{Serkowski} {et~al.}(1975){Serkowski}, {Mathewson}, \&
  {Ford}}]{serkowski1975}
{Serkowski}, K., {Mathewson}, D.~S., \& {Ford}, V.~L. 1975, \apj, 196, 261

\bibitem[{{Showman} \& {Kaspi}(2013)}]{showman2013}
{Showman}, A.~P., \& {Kaspi}, Y. 2013, \apj, 776, 85

\bibitem[{{Stefansson} {et~al.}(2017){Stefansson}, {Mahadevan}, {Hebb},
  {Wisniewski}, {Huehnerhoff}, {Morris}, {Halverson}, {Zhao}, {Wright},
  {O'rourke}, {Knutson}, {Hawley}, {Kanodia}, {Li}, {Hagen}, {Liu}, {Beatty},
  {Bender}, {Robertson}, {Dembicky}, {Gray}, {Ketzeback}, {McMillan}, \&
  {Rudyk}}]{stefansson2017}
{Stefansson}, G., {Mahadevan}, S., {Hebb}, L., {et~al.} 2017, \apj, 848, 9

\bibitem[{{Stetson}(1987)}]{Stetson1987}
{Stetson}, P.~B. 1987, \pasp, 99, 191

\bibitem[{{Stolker} {et~al.}(2017){Stolker}, {Min}, {Stam}, {Molli{\`e}re},
  {Dominik}, \& {Waters}}]{Stolker2017}
{Stolker}, T., {Min}, M., {Stam}, D.~M., {et~al.} 2017, \aap, 607, A42

\bibitem[{{Tan} \& {Showman}(2017)}]{tan2017}
{Tan}, X., \& {Showman}, A.~P. 2017, \apj, 835, 186

\bibitem[{{The Astropy Collaboration} {et~al.}(2018){The Astropy
  Collaboration}, {Price-Whelan}, {Sip{\H o}cz}, {G{\"u}nther}, {Lim},
  {Crawford}, {Conseil}, {Shupe}, {Craig}, {Dencheva}, {Ginsburg},
  {VanderPlas}, {Bradley}, {P{\'e}rez-Su{\'a}rez}, {de Val-Borro}, {Aldcroft},
  {Cruz}, {Robitaille}, {Tollerud}, {Ardelean}, {Babej}, {Bachetti}, {Bakanov},
  {Bamford}, {Barentsen}, {Barmby}, {Baumbach}, {Berry}, {Biscani}, {Boquien},
  {Bostroem}, {Bouma}, {Brammer}, {Bray}, {Breytenbach}, {Buddelmeijer},
  {Burke}, {Calderone}, {Cano Rodr{\'{\i}}guez}, {Cara}, {Cardoso},
  {Cheedella}, {Copin}, {Crichton}, {D{\'A}vella}, {Deil}, {Depagne},
  {Dietrich}, {Donath}, {Droettboom}, {Earl}, {Erben}, {Fabbro}, {Ferreira},
  {Finethy}, {Fox}, {Garrison}, {Gibbons}, {Goldstein}, {Gommers}, {Greco},
  {Greenfield}, {Groener}, {Grollier}, {Hagen}, {Hirst}, {Homeier}, {Horton},
  {Hosseinzadeh}, {Hu}, {Hunkeler}, {Ivezi{\'c}}, {Jain}, {Jenness}, {Kanarek},
  {Kendrew}, {Kern}, {Kerzendorf}, {Khvalko}, {King}, {Kirkby}, {Kulkarni},
  {Kumar}, {Lee}, {Lenz}, {Littlefair}, {Ma}, {Macleod}, {Mastropietro},
  {McCully}, {Montagnac}, {Morris}, {Mueller}, {Mumford}, {Muna}, {Murphy},
  {Nelson}, {Nguyen}, {Ninan}, {N{\"o}the}, {Ogaz}, {Oh}, {Parejko}, {Parley},
  {Pascual}, {Patil}, {Patil}, {Plunkett}, {Prochaska}, {Rastogi}, {Reddy
  Janga}, {Sabater}, {Sakurikar}, {Seifert}, {Sherbert}, {Sherwood-Taylor},
  {Shih}, {Sick}, {Silbiger}, {Singanamalla}, {Singer}, {Sladen}, {Sooley},
  {Sornarajah}, {Streicher}, {Teuben}, {Thomas}, {Tremblay}, {Turner},
  {Terr{\'o}n}, {van Kerkwijk}, {de la Vega}, {Watkins}, {Weaver}, {Whitmore},
  {Woillez}, \& {Zabalza}}]{astropy2018}
{The Astropy Collaboration}, {Price-Whelan}, A.~M., {Sip{\H o}cz}, B.~M.,
  {et~al.} 2018, ArXiv e-prints, arXiv:1801.02634

\bibitem[{{Vacca} {et~al.}(2003){Vacca}, {Cushing}, \& {Rayner}}]{vacca2003}
{Vacca}, W.~D., {Cushing}, M.~C., \& {Rayner}, J.~T. 2003, \pasp, 115, 389

\bibitem[{{Wardle} \& {Kronberg}(1974)}]{wardle1974}
{Wardle}, J.~F.~C., \& {Kronberg}, P.~P. 1974, \apj, 194, 249

\bibitem[{{Whittet} {et~al.}(1992){Whittet}, {Martin}, {Hough}, {Rouse},
  {Bailey}, \& {Axon}}]{whittet1992}
{Whittet}, D.~C.~B., {Martin}, P.~G., {Hough}, J.~H., {et~al.} 1992, \apj, 386,
  562

\bibitem[{{Wilson} {et~al.}(2003){Wilson}, {Eikenberry}, {Henderson},
  {Hayward}, {Carson}, {Pirger}, {Barry}, {Brandl}, {Houck}, {Fitzgerald}, \&
  {Stolberg}}]{wilson2003}
{Wilson}, J.~C., {Eikenberry}, S.~S., {Henderson}, C.~P., {et~al.} 2003, in
  \procspie, Vol. 4841, Instrument Design and Performance for Optical/Infrared
  Ground-based Telescopes, ed. M.~{Iye} \& A.~F.~M. {Moorwood}, 451--458

\bibitem[{Zhang \& Showman(2014)}]{Zhang2014}
Zhang, X., \& Showman, A.~P. 2014, \apjl, 788, L6

\end{thebibliography}

\end{document}